\documentclass[]{JHEP3}

\def\simlt{\stackrel{<}{{}_\sim}}
\def\simgt{\stackrel{>}{{}_\sim}}
\def\beq{\begin{equation}}
\def\eeq{\end{equation}}
\def\bea{\begin{eqnarray}}
\def\eea{\end{eqnarray}}

\title{Constructing the large mixing angle MNS matrix in see-saw models
with right-handed neutrino dominance}

\author{S. F. King\\
Department of Physics and Astronomy,
University of Southampton, Southampton, SO17 1BJ, U.K.}


\abstract
{Recent SNO results strongly favour the large mixing angle (LMA) 
MSW solar solution. We argue that 
there are only two technically natural low energy neutrino mass
matrix structures consistent with the LMA MSW solution, 
corresponding to either a hierarchy or an inverted
hierarchy with pseudo-Dirac neutrinos. We construct the
MNS matrix to leading order in the small angle $\theta_{13}$
including the neutrino {\em and} charged lepton mixing
angles and phases, the latter playing a crucial r\^{o}le for allowing
the inverted hierarchy case to be consistent with the LMA MSW solution.
We then consider the see-saw mechanism with 
right-handed neutrino dominance and show how the successful neutrino
mass matrix structures may be constructed with no tuning and with
small radiative corrections, leading to a full, partial or inverted
neutrino mass hierarchy. In each case we derive approximate analytic relations
between the input see-saw parameters and the resulting neutrino
masses, mixing angles and phases, which will provide a useful guide
for unified model building. 
For the hierarchical cases the LMA MSW solution gives a soft lower bound
$|U_{e3}|\simgt 0.1$, just below the current CHOOZ limit.
Both hierarchical and inverted hierarchical cases predict small
$\beta \beta_{0\nu}$ with $|m_{ee}|\sim 0.007$ eV within
the sensitivity of future proposals such as GENIUS.
Successful leptogenesis is possible if the dominant right-handed
neutrino is the heaviest one, but the leptogenesis phase is
unrelated to the MNS phases.}

\preprint{SHEP 02-09 \\ hep-ph/0204360}

\begin{document}

\section{Introduction}

Recent SNO results on the neutral current (NC) flux \cite{Ahmad:2002jz}
and the day-night effects \cite{Ahmad:2002ka}, when combined
with other solar neutrino data especially that of Super-Kamiokande
\cite{Fukuda:2001nk}
strongly favour the large mixing angle (LMA) MSW solar solution \cite{MSW}
with three active light neutrino states, and
$\theta_{12} \approx \pi/6$, 
$\Delta m_{21}^2\approx 5\times 10^{-5}{\rm eV}^2$
\cite{Barger:2002iv}, \cite{Bandyopadhyay:2002xj}, \cite{Bahcall:2002}.
The atmospheric neutrino data is consistent with
maximal $\nu_{\mu}- \nu_{\tau}$ neutrino mixing \cite{Fukuda:1998mi}
with $|\Delta m_{32}^2|\approx 2.5\times 10^{-3}{\rm eV}^2$
and the sign of $\Delta m_{32}^2$ undetermined. 
The CHOOZ experiment limits $\theta_{13} \simlt 0.2$
over the favoured atmospheric range \cite{Apollonio:1999ae}.
The combined neutrino data is well described by an MNS matrix 
\cite{Maki:1962mu} with 
$\theta_{23} \approx \pi/4$, $\theta_{12} \approx \pi/6$, 
$\theta_{13} \simlt 0.2$, which we refer to as the LMA MNS matrix.
It is clear that neutrino oscillations, which 
only depend on $\Delta m_{ij}^2\equiv m_i^2-m_j^2$, 
gives no information about the absolute value of the neutrino mass
eigenvalues $m_i$. Recent results from the 2df galaxy redshift survey
indicate that $\sum m_i <1.8 {\rm eV} (95\%C.L.)$
under certain mild assumptions \cite{Elgaroy:2002bi}.
Combined with the solar and atmospheric oscillation data
this brackets the heaviest neutrino mass to be
in the approximate range 0.04-0.6 eV. The fact that the mass of the
heaviest neutrino is known to within an order of magnitude represents
remarkable progress in neutrino physics over recent years.

The basic possible
patterns of neutrino mass consistent with this data are:
(i) hierarchy (full $m_1\ll m_2\ll m_3$, or
partial $m_1 \simlt  m_2 \ll m_3$), 
(ii) inverted hierarchy ($m_1\approx m_2 \gg m_3$, or the pseudo-Dirac form
$-m_1\approx m_2 \gg m_3$),  
or (iii) degenerate $m_1^2\approx m_2^2\approx m_3^2$
\cite{SelModels}.
Although oscillation data does not distinguish between these
possibilities, the theoretical requirement that the
neutrino spectrum is generated in a technically natural way,
does provide an additional guiding principle. 
For example it is clear that a
degenerate mass scale $m^2\gg |\Delta m_{ij}^2|$ implies
very small fractional neutrino mass splittings.
For example, if $m_3=0.500 {\rm eV}$ 
then atmospheric oscillations require 
neutrino masses $m_2=0.497 {\rm eV}$.
The problem is one of tuning, both to set up
the small mass splitting at the high energy scale, and to preserve
it in the presence of radiative corrections
\cite{Ellis:1999my}.
In general such a neutrino spectrum is not technically natural, 
since small perturbations in the high energy input parameters
will violate the low energy degeneracy.
One can envisage a technically
natural mechanism which would lead to a degenerate pair of neutrinos
with opposite sign masses $m_1\approx -m_2$, but to achieve the full
three-fold degeneracy is much more difficult
\cite{Barbieri:1999km},\cite{Chankowski:2001fp}.
A similar
objection can be raised against the inverted
hierarchical spectrum in which the almost degenerate neutrinos
have the same sign masses $m_1\approx m_2$ where  
for $m_1=0.0500 {\rm eV}$ we require $m_2=0.0497 {\rm eV}$
for the LMA MSW solution.
In these kinds of inverted hierarchy models there is no natural
mechanism that can set and preserve such mass splittings, however as
we shall see, in the inverted hierarchy model with 
a pseudo-Dirac neutrino pair corresponding to opposite
sign masses $m_1\approx -m_2$, a natural mechanism is possible.

Assuming the LMA MNS matrix plus 
naturalness implies that there are then just two leading order forms for the
light physical effective Majorana 
neutrino mass matrix $m^{\nu}_{LL}$ (where LL means that
it couples left-handed neutrinos to left-handed neutrinos)
corresponding to
either a hierarchical spectrum with $m\approx m_3\gg m_1,m_2$
arising from:
\begin{equation}
m^{\nu}_{LL}\approx 
\left( \begin{array}{ccc}
0 & 0 & 0    \\
0 & 1 & 1 \\
0 & 1 & 1
\end{array}
\right)\frac{m}{2} 
\label{H}
\end{equation}
or an inverted hierarchical spectrum with a pseudo-Dirac pair of
neutrinos with $m\approx m_2\approx -m_1\gg m_3$
arising from:
\begin{equation}
m^{\nu}_{LL}\approx 
\left( \begin{array}{rrr}
0 & 1 & -1    \\
1 & 0 & 0 \\
-1 & 0 & 0
\end{array}
\right)\frac{m}{\sqrt{2}} 
\label{I}
\end{equation}
where in both cases 
$m\approx \sqrt{|\Delta m_{32}^2|}\approx 5\times 10^{-2}{\rm eV}$.
It is remarkable that only by assuming the LMA MNS matrix plus
naturalness arguments we are led to only two possible leading order
structures for the neutrino mass matrix, both of which
have a zero in the 11 element corresponding to a small
$\beta \beta_{0\nu}$ rate corresponding to $|m_{ee}|=0$ at leading order.

For the hierarchical case the experimental requirement
$|\Delta m_{32}^2|\gg |\Delta m_{21}^2|$ implies
that $m_3^2\gg m_2^2$ which again looks technically unnatural
since we would expect two roughly equal eigenvalues once the
lower block is diagonalised. 
For the LMA MSW solution we only require a mild hierarchy,
$|m_2|/|m_3|\sim 0.1$, 
and this will require accidental cancellations of order 10\% 
to take place in the diagonalisation of the lower block
of the neutrino mass matrix. Although the degree of tuning
necessary to achieve the hierarchy is not large, it does imply that
the radiative corrections of the
matrix elements will be competetive with the amount by which they 
need to be tuned, and hence that
radiative corrections will be very important in
determining the low energy spectrum. In the case of
see-saw models \cite{seesaw} the radiative
corrections may be sufficient to destroy (or create) 
the cancellations necessary to achieve the desired hierarchy
\cite{Babu:1993qv},\cite{Ellis:1999nk}.
\footnote{The one-loop beta function coefficients in  
the Standard Model have recently been corrected in \cite{Antusch:2001ck},
\cite{Antusch:2001vn}. The two loop result for the MSSM is given in
\cite{Antusch:2002ek}.}
In all cases radiative corrections will be important and the
neutrino masses and mixings calculated in a given high energy unified
theory will not be simply related to the low energy ones.

The above situation could be improved
if it were possible to make the neutrino mass hierarchy completely
natural. If the hierarchy $m_3\gg m_2$ could 
emerge without any tuning at all, not even at the level of
10\% accidental cancellations, then the low energy spectrum would
faithfully preserve the nature of the spectrum calculated at high energy, 
without being severely affected by radiative
corrections. If this could be achieved then
the low energy measurements would provide a direct
window into the nature of the high energy theory.
Therefore it is interesting to ask 
whether it is possible for the neutrino mass
hierarchy to arise in a completely natural way? Indeed this is possible
if the see-saw mechanism \cite{seesaw} is supplemented by
a mechanism known as single right-handed neutrino dominance
\cite{King:1998jw}, \cite{King:1999cm}, \cite{King:2000mb}.
According to single right-handed neutrino 
dominance one of the right-handed neutrinos
makes the dominant contribution to the lower block of 
$m^{\nu}_{LL}$ causing its determinant to approximately vanish,
and thereby leading to $|m_2|\ll |m_3|$ without relying on
accidental cancellations which are subject to important radiative
corrections. Single right-handed neutrino dominance does not mean
that there is only a single right-handed neutrino, only that one of
the right-handed neutrinos is making the dominant contribution.
If the dominant right-handed neutrino is denoted $\nu_{R3}$
with a heavy Majorana mass $Y$ and Dirac couplings $d,e,f$
to the weak eigenstate neutrinos $\nu_{e},\nu_{\mu},\nu_{\tau}$
given by $\nu_{R3}(d\nu_{e}+e\nu_{\mu}+f\nu_{\tau})$
then according to the see-saw mechanism this will result
in a light physical neutrino $\nu_3\approx d\nu_{e}+e\nu_{\mu}+f\nu_{\tau}$
of mass $m_3\approx (d^2+e^2+f^2)/Y$  \cite{King:1998jw}
together with two light orthogonal combinations of neutrinos
which would be massless in the limit that there is only
a single right-handed neutrino. 
The requirements of a maximal atmospheric angle $\theta_{23}\approx \pi/4$
and a small CHOOZ angle $\theta_{13}\ll 1$ imply the relation
$d\ll e\approx f$ \cite{King:1998jw}.

In order to account for the solar data we must consider the effect of 
the sub-leading right-handed neutrinos. These perturb the spectrum
leading to a small second neutrino mass $m_2^2\ll m_3^2$. 
The strength of the hierarchy
is controlled by the relative importance of the 
sub-leading right-handed neutrinos, rather than relying on accidental
cancellations. If the sub-leading contribution is dominated by
a single sub-leading right-handed neutrino $\nu_{R2}$ with mass $X$
and couplings $\nu_{R2}(a\nu_{e}+b\nu_{\mu}+c\nu_{\tau})$, then
this leads to a second neutrino mass of order 
$m_2 \sim (1/2)(b-c)^2/X$ which only depends on the subleading
parameters because we must have $m_2=0$ in the limit of there being only a 
single right-handed neutrino \cite{King:2000mb}.
The solar angle was given by 
$\tan \theta_{12}\sim  \sqrt{2}a/(b-c)$ \cite{King:2000mb}.
Note that the solar angle is
completely determined by the sub-leading couplings,
due to a natural cancellation of the leading contributions \cite{King:2000mb}.
The lightest neutrino mass $m_1$ is generated by the sub-sub-leading
couplings due to the right-handed neutrino
$\nu_{R1}$ with mass $X'$ leading
to a full neutrino mass hierarchy $m_1\ll m_2 \ll m_3$ 
\cite{King:2000mb}.
We shall refer to this as sequential sub-dominance.

The sub-leading contribution may alternatively result from two equally
contributing sub-dominant right-handed neutrinos in which case 
only a partial neutrino mass hierarchy results $m_1\simlt m_2 \ll m_3$
and the results above will be different \cite{King:2000mb}.
While the full hierarchy results from an approximately
diagonal right-handed neutrino mass matrix, a partial
hierarchy may result from three possible textures
for the right-handed neutrino mass matrix 
namely diagonal, democratic or off-diagonal 
where the nomenclature refers to the upper block \cite{King:1999cm}:
\begin{equation}
M_{RR}^{\rm diag}=
\left( \begin{array}{ccc}
X' & 0 & 0    \\
0 & X & 0 \\
0 & 0 & Y
\end{array}
\right) 
\label{diag}
\end{equation}
\begin{equation}
M_{RR}^{\rm dem}=
\left( \begin{array}{ccc}
X & X & 0    \\
X & X & 0 \\
0 & 0 & Y
\end{array}
\right) 
\label{dem}
\end{equation}
\begin{equation}
M_{RR}^{\rm off-diag}=
\left( \begin{array}{ccc}
0 & X & 0    \\
X & 0 & 0 \\
0 & 0 & Y
\end{array}
\right) 
\label{off-diag}
\end{equation}
where the heavy Majorana neutrino mass matrices
$M_{RR}$ couple right-handed neutrinos to right-handed neutrinos.
Note that though the right-handed neutrinos in Eq.\ref{off-diag}
have a common pseudo-Dirac mass $X$, this may or may not
lead to the lighter two physical neutrinos arising from the
matrix in Eq.\ref{H} having a pseudo-Dirac mass.
As discussed \cite{King:2000mb} Eq.\ref{diag} can lead to either a
full or a partial hierarchy, while Eqs.\ref{dem} and \ref{off-diag}
only lead to partial hierarchies.
We shall refer to the cases where partial hierarchies result
from the structures in Eqs.\ref{diag},\ref{dem} and \ref{off-diag}
as diagonal sub-dominance,
democratic sub-dominance and off-diagonal sub-dominance.
We shall continue to refer the case where a full hierarchy
results from the diagonal structure in Eq.\ref{diag} 
as sequential sub-dominance.
Clearly they are all sub-classes of single right-handed
neutrino dominance since in all cases the dominant
contribution to the neutrino mass matrix comes from
the single right-handed neutrino with mass $Y$. 
The radiative corrections in such an approach would be expected to
be small since the hierarchy is now completely natural, and this
has been verified explicitly by a study of each of the cases
in Eqs.\ref{diag},\ref{dem},\ref{off-diag} the result of which shows that
radiative corrections typically change the physical predictions by only
a few per cent \cite{King:2000hk}.

The neutrino mass matrix in Eq.\ref{I} corresponding to an inverted
hierarchy with opposite sign masses $m_1\approx -m_2$ and 
$-m_1\approx m_2\gg m_3$, can be reproduced by three
right-handed neutrinos with the texture in Eq.\ref{off-diag}
\cite{King:2001ce}. However an important difference is that now 
it is the off-diagonal pair of right-handed neutrinos with
pseudo-Dirac mass $X$ that dominates the neutrino mass matrix 
leading to the pseudo-Dirac structure of neutrino masses
in Eq.\ref{I}, with the right-handed neutrino of mass $Y$ now giving
the sub-dominant contributions \cite{King:2001ce}.
We shall refer to this as off-diagonal right-handed neutrino dominance.
As in the hierarchical cases based on single right-handed neutrino
dominance, the resulting inverted hierarchical spectrum does not
rely on any accidental cancellations and is technically natural.
The radiative corrections arising to the inverted neutrino mass
spectrum arising from off-diagonal right-handed neutrino dominance
have been studied and shown to be only a few per cent \cite{King:2001ce}.

In this paper we consider
neutrino mass matrices with the leading order structures in
Eqs.\ref{H},\ref{I}, and diagonalise each of them
to leading order in $\theta_{13}$ to extract
the neutrino masses, mixing angles and phases. The
MNS matrix is then constructed
to leading order in the small angle $\theta_{13}$
including the neutrino {\em and} charged lepton mixing
angles and phases, the latter playing a crucial r\^{o}le for allowing
the inverted hierarchy solution to be consistent with the LMA MSW solution.
We then go on to show how the neutrino mass matrix structures
in Eqs.\ref{H},\ref{I}
may be constructed naturally from the see-saw mechanism with 
right-handed neutrino dominance, with no tuning and with
small radiative corrections, leading to a full, partial or inverted
neutrino mass hierarchy. In each case we derive approximate analytic relations
between the input see-saw mass matrices and the resulting neutrino
masses, mixing angles and phases. The goal of this analysis is to
provide a useful and reliable guide for constructing 
the LMA MNS matrix in unified models \cite{unified}.

The analysis builds on that
in \cite{King:2000mb} and \cite{King:2001ce}, by
including the effects of phases and the charged lepton mixing angles,
which we did not previously consider.
It is sufficient in a top-down approach to work to order
$\theta_{13}$, since although the radiative corrections are
only a few per cent (due to right-handed neutrino dominance),
this is sufficient to wash out the order $\theta_{13}^2$ corrections.
\footnote{By contrast in a bottom-up approach the 
order $\theta_{13}^2$ corrections may be considered
\cite{Lavignac:2002gf}. However the effect of charged lepton mixing 
angles was not considered in \cite{Lavignac:2002gf}
and this can significantly affect the
conclusions based on naturalness arguments.}
The leading order results which we present here give the
simple relations between the see-saw parameters necessary in order
to obtain the phenomenologically successful LMA MNS matrix.
Ultimately the masses and mixing angles
in unified models must be calculated numerically, including the 
radiative corrections. The purpose of the analytic results we present here
is to provide insight into the construction of unified models
which must then be studied numerically.

The paper is organised as follows.
In section 2 we review the construction of 
the MNS matrix starting from general
complex neutrino and charged lepton mass matrices,
with equivalence of different parametrisations of the MNS matrix
discussed in appendix \ref{equivalence}. 
We also discuss the charged lepton contributions
to the MNS matrix, where the natural expectation is that the
charged lepton mixing angles are all small, and give
an expansion of the MNS matrix to leading order in the small angles,
with details relegated to appendix \ref{charged}.
In section 3 we construct the successful neutrino mass matrix
structures in a natural way using the see-saw mechanism with
right-handed neutrino dominance.
In section 3.1 we derive results for
single right-handed neutrino dominance with 
sequential sub-dominance, corresponding to a full neutrino mass hierarchy.
In section 3.2 we derive results for
single right-handed neutrino dominance with off-diagonal
sub-dominance, corresponding to a partial neutrino mass hierarchy.
In section 3.3 we derive results for
off-diagonal right-handed neutrino dominance, corresponding to 
an inverted neutrino mass hierarchy, which we show is consistent
with the LMA MSW solution once the effect of charged lepton mixing
angles is taken into account.
For each type of right-handed
neutrino dominance we derive useful analytic expressions for
neutrino masses and mixing angles in terms of the see-saw mass
matrices, and show that for the hierarchical cases
the LMA MSW solution gives a soft lower bound
$|U_{e3}|\simgt 0.1$, just below the curent CHOOZ limit.
The results obtained in section 3 rely on analytic
diagonalisation methods discussed in appendices \ref{proceedure},
\ref{hierarchical}, \ref{inverted}.
In section 4 we give two physical application of the results.
In section 4.1 we consider $\beta \beta_{0\nu}$, and show that
both hierchical and inverted hierarchical cases predict small
$\beta \beta_{0\nu}$ with $|m_{ee}|\sim 0.007$ eV within
the sensitivity of future proposals such as GENIUS.
In section 4.2 we discuss leptogenesis, and show that
successful leptogenesis is possible if the dominant right-handed
neutrino is the heaviest one, but the leptogenesis phase is
unrelated to the MNS phases.
Section 5 concludes the paper.

\section{Constructing the MNS matrix}

The charged lepton masses and the neutrino masses
are given by the eigenvalues
of the complex charged lepton mass matrix
$m^E_{LR}$ and the complex symmetric neutrino Majorana matrix
$m_{LL}^{\nu}$, obtained by diagonalising these mass
matrices,
\beq
V^{E_L}m^E_{LR}{V^{E_R}}^{\dagger}=
\left( \begin{array}{ccc}
m_e & 0 & 0    \\
0 & m_{\mu} & 0 \\
0 & 0 & m_{\tau}
\end{array}
\right) 
\label{diag1}
\eeq
\beq
V^{\nu_L}m_{LL}^{\nu}{V^{\nu_L}}^T=
\left( \begin{array}{ccc}
m_1 & 0 & 0    \\
0 & m_2 & 0 \\
0 & 0 & m_3
\end{array}
\right) 
\label{diag2}
\eeq
where $V^{E_L}$, $V^{E_R}$, $V^{\nu_L}$ are unitary tranformations 
on the left-handed charged lepton fields $E_L$, right-handed charged
lepton fields $E_R$, and left-handed neutrino fields $\nu_L$
which put the mass matrices into diagonal form with real
eigenvalues. 

The MNS matrix is then constructed by
\beq
U_{MNS}=V^{E_L}{V^{\nu_L}}^{\dagger}
\label{MNS}
\eeq
In appendix \ref{equivalence} we discuss the relationship between
different parametrisations of the MNS matrix. In this paper
we find it convenient to parametrise the MNS matrix as
\beq
U_{MNS}=U_{23}U_{13}U_{12}
\label{MNS2}
\eeq
which involves just three irremoveable physical phases $\delta_{ij}$.
In this parametrisation the Dirac phase $\delta$ 
which enters the CP odd part of
neutrino oscillation probabilities is given by
\beq
\delta = \delta_{13}-\delta_{23}-\delta_{12}.
\label{Dirac}
\eeq
where the phases are defined in appendix \ref{equivalence}.

The MNS matrix is constructed in Eq.\ref{MNS} as a product
of a unitary matrix from the charged lepton sector $V^{E_L}$
and a unitary matrix from the neutrino sector ${V^{\nu_L}}^{\dagger}$.
Each of these unitary matrices may be parametrised by the
parametrisation of $V^{\dagger}$ in Eq.\ref{V1}.
Thus we write
\beq
{V^{\nu_L}}^{\dagger}
=P_2^{\nu_L}R_{23}^{\nu_L}R_{13}^{\nu_L}P_1^{\nu_L}R_{12}^{\nu_L}P_3^{\nu_L}
\label{VnuL}
\eeq
\beq
{V^{E_L}}^{\dagger}
=P_2^{E_L}R_{23}^{E_L}R_{13}^{E_L}P_1^{E_L}R_{12}^{E_L}P_3^{E_L}
\label{VEL}
\eeq
where the Euler angles and phases are defined as in 
Eqs.\ref{R23}-\ref{P3} 
but now there are independent angles
and phases for the left-handed neutrino and charged lepton sectors
distinguished by the superscripts $\nu_L$ and $E_L$.
As shown in appendix \ref{charged} the MNS matrix can be expanded
in terms of neutrino and charged lepton mixing angles and phases
to leading order in the charged lepton mixing angles which we
argue must be small. 
By comparing Eq.\ref{MNS7} to Eq.\ref{MNS6} we have:
\bea
s_{23}e^{-i\delta_{23}}
& \approx &
s_{23}^{\nu_L}e^{-i\delta_{23}^{\nu_L}}
-\theta_{23}^{E_L} 
c_{23}^{\nu_L}e^{-i\delta_{23}^{E_L}}
\label{chlep23}
\\
\theta_{13}e^{-i\delta_{13}}
& \approx &
\theta_{13}^{\nu_L}e^{-i\delta_{13}^{\nu_L}}
-\theta_{13}^{E_L}c_{23}^{\nu_L}e^{-i\delta_{13}^{E_L}}
+\theta_{12}^{E_L}s_{23}^{\nu_L}e^{i(-\delta_{23}^{\nu_L}-\delta_{12}^{E_L})}
\label{chlep13}
\\
s_{12}e^{-i\delta_{12}} 
& \approx &
s_{12}^{\nu_L}e^{-i\delta_{12}^{\nu_L}}
+\theta_{23}^{E_L}s_{12}^{\nu_L}e^{-i\delta_{12}^{\nu_L}}
+\theta_{13}^{E_L}
c_{12}^{\nu_L}s_{23}^{\nu_L}e^{i(\delta_{23}^{\nu_L}-\delta_{13}^{E_L})}
-\theta_{12}^{E_L}
c_{23}^{\nu_L}c_{12}^{\nu_L}e^{-i\delta_{12}^{E_L}}
\label{chlep12}
\eea
Clearly the large atmospheric and solar angles $\theta_{23}$ and $\theta_{12}$
arise mainly from the neutrino sector with small corrections
from the charged lepton angles. However $\theta_{13}$
receives important contributions not just from $\theta_{13}^{\nu_L}$,
but also from the charged lepton angles
$\theta_{12}^{E_L}$, and $\theta_{13}^{E_L}$.
In models where $\theta_{13}^{\nu_L}$ is
extremely small, $\theta_{13}$ may originate almost entirely from 
the charged lepton sector, since for example $\theta_{12}^{E_L}$
may be roughly equal to the Cabibbo angle in some models.
Charged lepton contributions could also be important in models
where $\theta_{12}^{\nu_L}=\pi /4$ very accurately, since
corrections from the charged lepton mixing angles
may allow consistency with the LMA MSW solution
which requires $\theta_{12}$ to be somewhat less than maximal.
Such effects will be important for the inverted hierarchy model,
for example, as we discuss in more detail later.

\section{See-saw models with right-handed neutrino dominance}
In this section we specialise to the case of the see-saw mechanism,
with right-handed neutrino dominance.
A particular high energy theory will involve a charged lepton 
Yukawa matrix $Y^E$, a neutrino Yukawa matrix $Y^{\nu}$
and a right-handed neutrino Majorana matrix $M_{RR}$.
The Higgs vacuum expectation values (vevs) may be absorbed in the
Yukawa matrices to give mass matrices $m^E_{LR}=v_1Y^E$ and 
$m_{LR}^{\nu}=v_2Y^{\nu}$, where we allow for two different 
vevs $v_1,v_2$ as is the case in supersymmetric models,
while in the standard model $v_1=v_2$, and LR means that
these are Dirac mass matrices in the left-right basis where the 
rows correspond to left-handed fields and the columns to right-handed
fields. At high energies these mass matrices $m^E_{LR}$,
$m_{LR}^{\nu}$ and $M_{RR}$ are ``born'' in a particular basis
which is defined by the particular unified theory.
In general $m^E_{LR}$ and $m_{LR}^{\nu}$ are general complex 
matrices, while $M_{RR}$ is complex symmetric.

In this ``theory'' basis the light effective Majorana neutrino mass
matrix is given, up to an overall irrelevant sign,
from the see-saw formula \cite{seesaw},\cite{Mohapatra:1980ia},
\cite{Schechter:1982cv}:
\beq
m_{LL}^{\nu}=m_{LR}^{\nu}M_{RR}^{-1}{m_{LR}^{\nu \ T}},
\label{seesaw}
\eeq
ignoring the effects of radiative corrections.

In order for our results to be maximally useful for top-down
model building we must work in the ``theory'' basis 
defined by the high energy theory.
In this basis all the mass matrices are in general off-diagonal, 
and in particular the charged lepton mass matrix will be off-diagonal,
although according to our naturalness arguments we would expect it
to yield small charged lepton mixing angles.
Right-handed neutrino dominance should be applied in this theory basis,
and questions of naturalness must be addressed within this
defining basis. However it may happen that in specific theories
certain simplifications naturally appear. For example the
right-handed neutrino mass matrix may take one
of the approximate forms in Eqs.\ref{diag}, \ref{dem}, \ref{off-diag}.

The general proceedure for diagonalising a general complex
matrix is discussed in appendix \ref{proceedure}.
The strategy is to diagonalise the charged lepton and neutrino
mass matrices $m^E_{LR}$ and $m_{LL}^{\nu}$ using,
\beq
{P_3^{E_L}}^{\ast}{R_{12}^{E_L}}^{T}{P_1^{E_L}}^{\ast}
{R_{13}^{E_L}}^{T}{R_{23}^{E_L}}^{T}{P_2^{E_L}}^{\ast}
m^E_{LR}
P_2^{E_R}R_{23}^{E_R}R_{13}^{E_R}P_1^{E_R}R_{12}^{E_R}P_3^{E_R}=
\left( \begin{array}{ccc}
m_e & 0 & 0    \\
0 & m_{\mu} & 0 \\
0 & 0 & m_{\tau}
\end{array}
\right) 
\label{diag3}
\eeq
\beq
{P_3^{\nu_L}}^{\ast}{R_{12}^{\nu_L}}^{T}{P_1^{\nu_L}}^{\ast}
{R_{13}^{\nu_L}}^{T}{R_{23}^{\nu_L}}^{T}{P_2^{\nu_L}}^{\ast}
m_{LL}^{\nu}
{P_2^{\nu_L}}^{\ast}{R_{23}^{\nu_L}}{R_{13}^{\nu_L}}
{P_1^{\nu_L}}^{\ast}{R_{12}^{\nu_L}}{P_3^{\nu_L}}^{\ast}
=
\left( \begin{array}{ccc}
m_1 & 0 & 0    \\
0 & m_2 & 0 \\
0 & 0 & m_3
\end{array}
\right) 
\label{diag4}
\eeq
The specific case of a hierarchical neutrino 
mass matrix is discussed in appendix \ref{hierarchical},
while the inverted hierarchical case is discussed
in appendix \ref{inverted}.

\subsection{Full hierarchy from sequential sub-dominance}
In this sub-section we shall consider the case where the
right-handed neutrino mass matrix takes an
approximately diagonal form as in Eq.\ref{diag}.
If it is approximately diagonal, then
it may easily be rotated to exactly diagonal form with real
eigenvalues, to give
\begin{equation}
M_{RR}^{\rm diag}=
\left( \begin{array}{ccc}
X' & 0 & 0    \\
0 & X & 0 \\
0 & 0 & Y
\end{array}
\right) 
\label{seq1}
\end{equation}
In the theory basis we shall write the complex Dirac mass matrix as
\begin{equation}
m^{\nu}_{LR}=
\left( \begin{array}{ccc}
a' & a & d    \\
b' & b & e \\
c' & c & f
\end{array}
\right) 
\label{dirac}
\end{equation}

Then using the see-saw formula \ref{seesaw} 
(valid for complex couplings) we find
the complex symmetric neutrino mass matrix,
\beq
m^{\nu}_{LL}
=
\left( \begin{array}{ccc}
\frac{a'^2}{X'}+\frac{a^2}{X}+\frac{d^2}{Y}
& \frac{a'b'}{X'}+\frac{ab}{X}+ \frac{de}{Y}
& \frac{a'c'}{X'}+\frac{ac}{X}+\frac{df}{Y}    \\
.
& \frac{b'^2}{X'}+\frac{b^2}{X}+\frac{e^2}{Y} 
& \frac{b'c'}{X'}+\frac{bc}{X}+\frac{ef}{Y}    \\
.
& .
& \frac{c'^2}{X'}+\frac{c^2}{X}+\frac{f^2}{Y} 
\end{array}
\right)
\label{seq2}
\eeq
The condition for single right-handed neutrino dominance with 
sequential sub-dominance was given as \cite{King:2000mb}
\beq
\frac{|e^2|,|f^2|,|ef|}{Y}\gg
\frac{|xy|}{X} \gg
\frac{|x'y'|}{X'}
\label{srhnd}
\eeq
where $x,y\in a,b,c$ and $x',y'\in a',b',c'$, and now we are dealing
with complex matrices we must consider the absolute value of the elements. 
For sequential subdominance we can essentially ignore the
contributions from the right-handed neutrino with mass $X'$, so
that the problem reduces to the simpler case of
two right-handed neutrinos \cite{King:2000mb} with
\beq
m^{\nu}_{LL}
\approx
\left( \begin{array}{ccc}
\frac{a^2}{X}+\frac{d^2}{Y}
& \frac{ab}{X}+ \frac{de}{Y}
& \frac{ac}{X}+\frac{df}{Y}    \\
.
& \frac{b^2}{X}+\frac{e^2}{Y} 
& \frac{bc}{X}+\frac{ef}{Y}    \\
.
& .
& \frac{c^2}{X}+\frac{f^2}{Y} 
\end{array}
\right)
\equiv 
\left( \begin{array}{ccc}
m^{\nu}_{11} & m^{\nu}_{12} & m^{\nu}_{13} \\
m^{\nu}_{12} & m^{\nu}_{22} & m^{\nu}_{23} \\
m^{\nu}_{13} & m^{\nu}_{23} & m^{\nu}_{33}
\end{array}
\right) 
\label{seq3}
\eeq
where we have made contact with the notation of Eq.\ref{mnu1}.
In order that Eq.\ref{seq3} resembles the leading order form of the
matrix in Eq.\ref{H} we require
\beq
|d|\ll |e|\approx |f|
\label{def}
\eeq
as first shown in \cite{King:1998jw}, although in the complex case
here we have taken the absolute values of the complex couplings.
It is straighforward to diagonalise the matrix in Eq.\ref{seq3}
by following the proceedure in appendix \ref{hierarchical}. 
According to Eq.\ref{mnu1} there is a complex phase $\phi^{\nu}_{ij}$
associated with each element of the mass matrix. In our previous results
the effect of phases was ignored \cite{King:2000mb}.
Here we include them, and discuss their effect on our previous
results. 

We first focus on the lower block
of Eq.\ref{seq3}, 
\beq
\left( \begin{array}{cc}
m^{\nu}_{22} & m^{\nu}_{23} \\
m^{\nu}_{23} & m^{\nu}_{33}
\end{array}
\right) 
\equiv
\left( \begin{array}{cc}
\frac{b^2}{X}+\frac{e^2}{Y} & \frac{bc}{X}+\frac{ef}{Y}    \\
\frac{bc}{X}+\frac{ef}{Y} & \frac{c^2}{X}+\frac{f^2}{Y} 
\end{array}
\right)
\label{seq5}
\eeq
This matrix may then be diagonalised by the $P_2^{\nu_L}$
re-phasing followed by
the 23 rotation 
\beq
\left( \begin{array}{cc}
\tilde{m}^{\nu}_{22} &  0\\
0 &  m_3'
\end{array}
\right) 
\equiv
{R^{\nu_L}_{23}}^T
\left( \begin{array}{cc}
m^{\nu}_{22}e^{-2i\phi^{\nu_L}_2}  
& m^{\nu}_{23}e^{-i(\phi^{\nu_L}_2+\phi^{\nu_L}_3)}  \\
m^{\nu}_{23}e^{-i(\phi^{\nu_L}_2+\phi^{\nu_L}_3)}
& m^{\nu}_{33}e^{-2i\phi^{\nu_L}_3} 
\end{array}
\right) 
R^{\nu_L}_{23}
\label{nustep2222}
\eeq
The 23 mixing angle may readily be
obtained from Eq.\ref{nu23L}, which is accurate to order $\theta_{13}$.
We shall find that in the hierarchical cases $\theta_{13}\simgt m_2/m_3$
and that this bound must be saturated for the LMA MSW solution.
Therefore in this case the results will be accurate to 
order $m_2/m_3$. From Eq.\ref{nu23L} we find
\beq
\tan 2\theta_{23}^{\nu_L}\approx
\frac{2e^{-i(\phi^{\nu}_{2}+\phi^{\nu}_{3})}\frac{ef}{Y}
(1+\epsilon_1-\epsilon_2)}
{(e^{-2i\phi^{\nu}_{3}}\frac{f^2}{Y}-e^{-2i\phi^{\nu}_{2}}\frac{e^2}{Y})}
\label{hi23}
\eeq
where
\beq
\epsilon_1 = \frac{\frac{bc}{X}}{\frac{ef}{Y}},\ \
\epsilon_2 = 
\frac{e^{-2i\phi^{\nu}_{3}}\frac{c^2}{X}
-e^{-2i\phi^{\nu}_{2}}\frac{b^2}{X}}
{e^{-2i\phi^{\nu}_{3}}\frac{f^2}{Y}-e^{-2i\phi^{\nu}_{2}}\frac{e^2}{Y}}
\label{eps}
\eeq
In the leading approximation, in which the corrections of order
$\epsilon_i$ are neglected, the phases arise solely from the complex
Dirac couplings $e=|e|e^{i\phi_e}$, $f=|f|e^{i\phi_f}$, and 
we have the elegant leading order result
\beq
\tan \theta_{23}^{\nu_L}\approx
\frac{|e|}{|f|}.
\label{seq123L}
\eeq
where the phases are fixed by
\beq
\phi^{\nu_L}_2-\phi^{\nu_L}_3=\phi_e-\phi_f
\label{phase5}
\eeq
This leading order result was first written down without phases in
\cite{King:1998jw}. The result in Eq.\ref{seq123L} demonstrates
that the phases on $e,f$ are not important in determining 
$\theta_{23}^{\nu_L}$ at leading order, only their absolute values
matter. 

It is instructive to take the determinant of both sides
of Eq.\ref{nustep2222},
\beq
\tilde{m}^{\nu}_{22}m_3'=
m^{\nu}_{22}e^{-2i\phi^{\nu_L}_2}  
m^{\nu}_{33}e^{-2i\phi^{\nu_L}_3} 
-
(m^{\nu}_{23})^2e^{-2i(\phi^{\nu_L}_2+\phi^{\nu_L}_3)}
=
e^{-2i(\phi^{\nu}_{2}+\phi^{\nu_L}_3)}\frac{(bf-ce)^2}{XY}
\label{det1}
\eeq
where the leading order terms proportional to $1/Y^2$ have cancelled
in constructing the determinant. This is no accident: it happens because 
the determinant vanishes in the limit of a single right-handed neutrino.
The fact that the determinant is small implies that
\beq
|m_3'|\gg |\tilde{m}^{\nu}_{22}|
\label{gg1}
\eeq
The natural origin of Eq.\ref{gg1}
is crucial both in obtaining a natural hierarchy $m_2\ll m_3$
and in obtaining a large solar angle. 
The naturally small 23 subdeterminant is the main consequence of single
right-handed neutrino dominance, as emphasised in \cite{King:1999cm}.

If we take the trace of both sides
of Eq.\ref{nustep2222}, using Eq.\ref{gg1},
then we may obtain the third neutrino mass to leading order 
\beq
m_3'\approx e^{i(2\phi^{\nu}_{e}-2\phi^{\nu_L}_2)}\frac{[|e|^2+|f|^2]}{Y}
\label{m3p2}
\eeq
\beq
m_3\approx \frac{[|e|^2+|f|^2]}{Y}
\eeq
A more accurate expression for $m'_3$ including the $m_2/m_3$
corrections can be readily obtained 
from Eq.\ref{m3p}, using Eqs.\ref{seq5},\ref{hi23}.
However the result is not very illuminating, and so is not worth
displaying explicitly, although it may be readily constructed if
required. The same comments apply to the remaining parameters,
which we only display explicitly to leading order.

The leading order result for $\tilde{m}^{\nu}_{22}$,
then follows from Eqs.\ref{m3p2}, \ref{det1},\ref{seq123L},\ref{phase5}
\beq
\tilde{m}^{\nu}_{22}
\approx
e^{-2i\phi^{\nu_L}_2}
\frac{(c_{23}^{\nu_L}b-s^{\nu_L}_{23}ce^{i(\phi^{\nu}_{e}-\phi^{\nu}_{f})})^2}{X}
\label{m22t}
\eeq
The leading order results for 
$\tilde{m}^{\nu}_{12}$, $\tilde{m}^{\nu}_{13}$ 
follow from Eq.\ref{nustep23}, which we write here as
\beq
\left( \begin{array}{c}
\tilde{m}^{\nu}_{12} \\
\tilde{m}^{\nu}_{13} 
\end{array}
\right) 
=
{R_{23}^{\nu_L}}^{T}
\left( \begin{array}{c}
m^{\nu}_{12}
e^{-i\phi^{\nu_L}_2} \\
m^{\nu}_{13}
e^{-i\phi^{\nu_L}_3}
\end{array}
\right) 
\label{nustep123}
\eeq
We find to leading order using Eqs.\ref{nustep123},
\ref{seq3},\ref{seq123L},\ref{phase5}
\beq
\tilde{m}^{\nu}_{12}
\approx
e^{-i\phi^{\nu_L}_2}\frac{a
(c_{23}^{\nu_L}b-s^{\nu_L}_{23}ce^{i(\phi^{\nu}_{e}-\phi^{\nu}_{f})})}{X}
\label{m12t}
\eeq
\beq
\tilde{m}^{\nu}_{13}
\approx
e^{-i\phi^{\nu_L}_2}
\left[
\frac{a(s^{\nu_L}_{23}b+c_{23}^{\nu_L}ce^{i(\phi^{\nu}_{e}-\phi^{\nu}_{f})})}
{X}
+e^{i\phi_e}
\frac{d\sqrt{|e|^2+|f|^2}}{Y}
\right]
\label{m13t}
\eeq
Note that the $1/Y$ terms have cancelled to leading order in
Eq.\ref{m12t} as in the real case \cite{King:2000mb}.

The 13 neutrino angle is given from Eq.\ref{nu13L}
\beq
\theta_{13}^{\nu_L}\approx
\frac{\tilde{m}^{\nu}_{13}}{m_3'}
\label{1nu13L}
\eeq
where the leading order form for
$\tilde{m}^{\nu}_{13}$ is given in Eq.\ref{m13t}, and
for $m_3'$ in Eq.\ref{m3p2}.
In the large $d$ limit
\beq
\frac{|de|}{Y},\frac{|df|}{Y}\gg \frac{|ab|}{X},\frac{|ac|}{X}
\label{larged}
\eeq
we find the simple leading order result \cite{King:2000mb},
\beq
\theta_{13}^{\nu_L}\approx \frac{|d|}{\sqrt{|e|^2+|f|^2}}
\label{2nu13L}
\eeq
with the phase $\phi^{\nu_L}_2$ fixed by
\beq
\phi^{\nu_L}_2\approx \phi_e-\phi_d
\eeq
Ignoring phases, this was the result previously quoted \cite{King:2000mb}.
On the other hand in the small $d$ limit
\beq
\frac{|de|}{Y},\frac{|df|}{Y}\ll \frac{|ab|}{X},\frac{|ac|}{X}
\eeq
we obtain the leading order result
\beq
\theta_{13}^{\nu_L}\approx\frac{1}{m_3'}e^{-i\phi^{\nu_L}_2}
\frac{a(s^{\nu_L}_{23}b+c_{23}^{\nu_L}ce^{i(\phi^{\nu}_{e}-\phi^{\nu}_{f})})}{X}
\label{3nu13L}
\eeq
where the phase $\phi^{\nu_L}_2$ is fixed by
Eq.\ref{absphase}.

After the 13 rotation the neutrino mass matrix is in block diagonal
form 
\beq
\left( \begin{array}{ccc}
\tilde{m}^{\nu}_{11} & \tilde{m}^{\nu}_{12} & 0 \\
\tilde{m}^{\nu}_{12} &  \tilde{m}^{\nu}_{22} &  0 \\
0 &  0 &  m_3'
\end{array}
\right) 
\label{1nustep3}
\eeq
where the leading order form for $\tilde{m}^{\nu}_{12}$ is given in 
Eq.\ref{m12t}, for $\tilde{m}^{\nu}_{22}$ in Eq.\ref{m22t}.
We need to find $\tilde{m}^{\nu}_{11}$ from Eq.\ref{num11}
\beq
\tilde{m}^{\nu_L}_{11}\approx m^{\nu_L}_{11}-
\frac{(\tilde{m}^{\nu_L}_{13})^2}{m_3'}
\label{1num11}
\eeq
Using the leading order form for $\tilde{m}^{\nu_L}_{13}$
in Eq.\ref{m13t}, and for $m_3'$ in Eq.\ref{m3p2},
and $\tilde{m}^{\nu_L}_{11}$ in Eq.\ref{seq3},
we find 
\beq
\tilde{m}^{\nu_L}_{11}\approx 
\frac{a^2}{X}-e^{-i\phi^{\nu}_{e}}\frac{2d}{\sqrt{|e|^2+|f|^2}}.
\frac{a(s^{\nu_L}_{23}b+c_{23}^{\nu_L}ce^{i(\phi^{\nu}_{e}-\phi^{\nu}_{f})})}{X}
\label{m11t}
\eeq
where the leading $1/Y$ dependence in $\tilde{m}^{\nu_L}_{11}$
has cancelled, as in the real case \cite{King:2000mb}.

In the leading order, valid up to corrections of order $m_2/m_3$,
we may ignore the second term in Eq.\ref{m11t} since this cannot
be larger than order $\theta_{13}^{\nu_L}$, as shown in Eq.\ref{2nu13L}.
Then the upper block which remains to be diagonalised in 
Eq.\ref{1nustep3} is given
from Eqs.\ref{m11t}, \ref{m12t}, \ref{m22t},
and after the phase transformations $P_1^{\nu_L}$ it is:
\beq
\left( \begin{array}{cc}
A^2 & AB \\
AB & B^2
\end{array}
\right) 
\label{22block}
\eeq
where
\beq
A=\frac{a}{\sqrt{X}}, \ \ B= e^{-i(\phi^{\nu_L}_2+\chi^{\nu_L})}
\frac{(c_{23}^{\nu_L}b-s^{\nu_L}_{23}ce^{i(\phi^{\nu}_{e}-\phi^{\nu}_{f})})}
{\sqrt{X}}
\label{AB}
\eeq
The leading order form in Eq.\ref{22block} clearly has a vanishing 
determinant 
\beq
m_1'm_2'\approx 0.
\label{det3}
\eeq
This is no surprise since
sequential sub-dominance corresponds approximately to the 
case of two right-handed neutrinos, and hence the lightest
neutrino mass is approximately zero, as in the case of a full 
hierarchy. The situation is analagous to the case of the lower block
with single right-handed neutrino dominance discussed above.
The second neutrino mass is simply given at leading order from the trace of
Eq.\ref{22block},
\beq
m_2'\approx A^2+B^2
\label{1m2p}
\eeq
The leading order 
12 neutrino mixing angle is given from Eq.\ref{nu12L},\ref{22block} 
\beq
\tan \theta_{12}^{\nu_L}\approx
\frac{A}{B}\approx 
\frac{ae^{i(\phi^{\nu_L}_2+\chi^{\nu_L})}}
{(c_{23}^{\nu_L}b-s^{\nu_L}_{23}ce^{i(\phi^{\nu}_{e}-\phi^{\nu}_{f})})}
\label{1nu12L}
\eeq
which is analagous to the real result in \cite{King:2000mb}.
The phase $\chi^{\nu_L}$ is fixed to give a real 12 angle as
in Eq.\ref{chi}
\beq
c_{23}^{\nu_L}|b|
\sin(\phi_b')
\approx
s^{\nu_L}_{23}|c|
\sin(\phi_c')
\label{chi1}
\eeq
and then
\beq
\tan \theta_{12}^{\nu_L}\approx
\frac{|a|}
{c_{23}^{\nu_L}|b|
\cos(\phi_b')-
s^{\nu_L}_{23}|c|
\cos(\phi_c')}
\label{2nu12L}
\eeq
where
\bea
\phi_b' &\equiv & \phi_b-\phi_a-\phi^{\nu}_{2}-\chi^{\nu_L},\\ 
\phi_c' &\equiv & \phi_c-\phi_a+\phi_e-\phi_f-\phi^{\nu}_{2}-\chi^{\nu_L}
\eea
Eq.\ref{2nu12L} shows that a large solar angle 
$\theta_{12}^{\nu_L}\sim \pi/6$ requires 
$|a| /(|b|\cos(\phi_b')-|c|\cos(\phi_c'))\sim 1$,
which is the complex analogue of the real condition
$\sqrt{2}a/(b-c)\sim 1$ derived in \cite{King:2000mb}.
The effect of phases is now quite important since the denominator
of Eq.\ref{2nu12L} may involve phase-dependent cancellations.
In the absence of such cancellations, the basic physical requirement
for a large mixing angle, namely that the sub-dominant Dirac coupling
$a$ be of order $b,c$ remains as in \cite{King:2000mb}.

From Eqs.\ref{m3p2},\ref{1m2p},\ref{det3}, we see that
sequential sub-dominance has generated a full neutrino mass hierarchy
\beq
m_1\ll m_2\ll m_3
\label{fullhi}
\eeq
and hence
\beq
\Delta m_{32}^2\approx m_3^2, \ \  \Delta m_{21}^2\approx m_2^2.
\label{delta1}
\eeq
Sequential sub-dominance naturally leads to the 
LMA MSW solution by assuming a mild hierarchy of couplings
in Eq.\ref{srhnd}.

We now discuss an order of magnitude lower bound on the 13 neutrino mixing
angle. Clearly such a bound can only be order of magnitude
since it is always possible to arrange vanishingly small values
of this angle by tuning parameters. However, in the absence of tuning,
it is possible to make some general statements about the expected
magnitude of this angle. The question is therefore, in the absence 
of unnatural cancellations, how small can the 13 neutrino angle be?
Clearly from Eqs.\ref{1nu13L},\ref{m13t}, in the absence of tuning,
the smallest values of $\theta_{13}^{\nu_L}$ correspond to setting
$d=0$, and hence the limit will be saturated by Eq.\ref{3nu13L}.
The 13 neutrino angle in Eq.\ref{3nu13L} cannot be made arbitrarily
small by setting $a=0$ since we have just seen that 
a large solar angle requires $a$ be of order $b,c$.
Therefore the numerator of Eq.\ref{3nu13L} is expected to be of a
similar order of magnitude to $m_2$ in Eq.\ref{1m2p},
and for a large solar angle we therefore obtain the order
of magnitude lower bound
\beq
|\theta_{13}^{\nu_L}|\simgt |m_2/m_3|
\label{13bound1}
\eeq
From Eq.\ref{MNS7}, since the charged lepton angles are all 
required to be small due to the naturalness arguments
$\theta_{12}^{E_L},\theta_{13}^{E_L} \simlt \theta_{13}$,
and using Eq.\ref{delta1},
we may elevate Eq.\ref{13bound1} to a bound on the MNS element,
\footnote{Similar bounds have been obtained previously
in somewhat different frameworks in \cite{Akhmedov:1999uw},
\cite{Feruglio:2002af},\cite{Lavignac:2002gf}}
\beq
|U_{e3}|^2\simgt |\Delta m_{21}^2/\Delta m_{32}^2|
\label{13bound2}
\eeq
The bound is roughly
proportional to $\tan \theta_{12}^{\nu_L}$, by eliminating $a$.
Interestingly Eq.\ref{13bound2} predicts that for the LMA solution
$|U_{e3}|$ should be just below the current CHOOZ current limit,
which raises the prospect that it could be measured at the
forthcoming long baseline (LBL) experiments.
We emphasise again, however, that the bound is a soft one
since it is based in the premise of there being no cancellations
in the construction of the MNS matrix, and so is really only
approximate to within an order of magnitude.

\subsection{Partial hierarchy from off-diagonal sub-dominance}

We now turn the case where at high energies the upper block of the
right-handed neutrino mass matrix has the approximate off-diagonal
form as in Eq.\ref{off-diag}. By small angle rotations it may then
be rotated into the form below with real mass parameters $X,Y$,
\begin{equation}
M_{RR}^{\rm off-diag}=
\left( \begin{array}{ccc}
0 & X & 0    \\
X & 0 & 0 \\
0 & 0 & Y
\end{array}
\right) 
\label{off1}
\end{equation}
Using the same notation for the Dirac couplings as in Eq.\ref{dirac},
the see-saw formula in Eq.\ref{seesaw} then gives in this case,
\beq
m^{\nu}_{LL}
=
\left( \begin{array}{ccc}
\frac{2aa'}{X}+\frac{d^2}{Y}
& \frac{a'b}{X}+\frac{ab'}{X}+ \frac{de}{Y}
& \frac{a'c}{X}+\frac{ac'}{X}+\frac{df}{Y}    \\
.
& \frac{2bb'}{X}+\frac{e^2}{Y} 
& \frac{b'c}{X}+\frac{bc'}{X}+\frac{ef}{Y}    \\
.
& .
& \frac{2cc'}{X}+\frac{f^2}{Y} 
\end{array}
\right)
\label{off2}
\eeq
The condition for single right-handed neutrino dominance with 
off-diagonal sub-dominance was given as \cite{King:2000mb}
\beq
\frac{|e^2|,|f^2|,|ef|}{Y}\gg
\frac{|xx'|}{X}
\label{srhnd2}
\eeq
where $x\in a,b,c$ and $x'\in a',b',c'$, and now we are dealing
with complex matrices we must consider the absolute value of the elements. 

As before, for the neutrino mass matrix to resemble the leading order
form in Eq.\ref{H} we also require the condition in Eq.\ref{def}.
Then the discussion of the leading order 23 neutrino mixing angle
follows exactly as in the sequential sub-dominance case,
with the lower block replaced by
\beq
\left( \begin{array}{cc}
m^{\nu}_{22} & m^{\nu}_{23} \\
m^{\nu}_{23} & m^{\nu}_{33}
\end{array}
\right) 
\equiv
\left( \begin{array}{cc}
\frac{2bb'}{X}+\frac{e^2}{Y} & \frac{b'c}{X}+\frac{bc'}{X}+\frac{ef}{Y}    \\
\frac{b'c}{X}+\frac{bc'}{X}+\frac{ef}{Y} & \frac{2cc'}{X}+\frac{f^2}{Y} 
\end{array}
\right)
\label{off3}
\eeq
From Eq.\ref{nu23L} we find the analagous result to Eq.\ref{hi23},
\beq
\tan 2\theta_{23}^{\nu_L}\approx
\frac{2e^{-i(\phi^{\nu}_{2}+\phi^{\nu}_{3})}\frac{ef}{Y}
(1+\epsilon'_1-\epsilon'_2)}
{(e^{-2i\phi^{\nu}_{3}}\frac{f^2}{Y}-e^{-2i\phi^{\nu}_{2}}\frac{e^2}{Y})}
\label{parhi23}
\eeq
where now
\beq
\epsilon'_1 = \frac{\frac{b'c}{X}+\frac{bc'}{X}}{\frac{ef}{Y}},\ \
\epsilon'_2 = 
\frac{e^{-2i\phi^{\nu}_{3}}\frac{2cc'}{X}
-e^{-2i\phi^{\nu}_{2}}\frac{2bb'}{X}}
{e^{-2i\phi^{\nu}_{3}}\frac{f^2}{Y}-e^{-2i\phi^{\nu}_{2}}\frac{e^2}{Y}}
\label{epsp}
\eeq
In the leading approximation, in which the corrections of order
$\epsilon_i$ are neglected, we obtain the same results as in 
Eqs.\ref{seq123L},\ref{phase5},
\beq
\tan \theta_{23}^{\nu_L}\approx
\frac{|e|}{|f|}.
\label{off4}
\eeq

The determinant of Eq.\ref{nustep2222} with Eq.\ref{off3} gives
\beq
\tilde{m}^{\nu}_{22}m_3'=
2e^{-2i(\phi^{\nu}_{2}+\phi^{\nu_L}_3)}\frac{(bf-ce)(b'f-c'e)}{XY}
\label{det2}
\eeq
where again the leading order terms in the determinant have cancelled
due to single right-handed neutrino dominance.
The trace of Eq.\ref{nustep2222} with Eq.\ref{off3} gives the same
leading order result for $m_3'$ is as in Eq.\ref{m3p2},
\beq
m_3'\approx e^{i(2\phi^{\nu}_{e}-2\phi^{\nu_L}_2)}\frac{[|e|^2+|f|^2]}{Y}
\label{m3p3}
\eeq
Hence from Eqs.\ref{det2}, \ref{m3p3}, 
\beq
\tilde{m}^{\nu}_{22}
\approx
2e^{-2i\phi^{\nu_L}_2}
\frac{(c_{23}^{\nu_L}b-s^{\nu_L}_{23}ce^{i(\phi^{\nu}_{e}-\phi^{\nu}_{f})})
(c_{23}^{\nu_L}b'-s^{\nu_L}_{23}c'e^{i(\phi^{\nu}_{e}-\phi^{\nu}_{f})})}
{X}
\label{m22t1}
\eeq
As before, henceforth we display only the leading order results explicitly,
with the corrections $m_2/m_3$ readily obtainable 
from our formalism if required.
The leading order results for 
$\tilde{m}^{\nu}_{12}$, $\tilde{m}^{\nu}_{13}$ 
follow from Eq.\ref{nustep123}.
We find to leading order using Eqs.\ref{nustep123},\ref{off2},
\beq
\tilde{m}^{\nu}_{12}
\approx
e^{-i\phi^{\nu_L}_2}
\left[
\frac{a'(c_{23}^{\nu_L}b-s^{\nu_L}_{23}ce^{i(\phi^{\nu}_{e}-\phi^{\nu}_{f})})
+a(c_{23}^{\nu_L}b'-s^{\nu_L}_{23}c'e^{i(\phi^{\nu}_{e}-\phi^{\nu}_{f})})}
{X}\right]
\label{1m12t}
\eeq
\beq
\tilde{m}^{\nu}_{13}
\approx
e^{-i\phi^{\nu_L}_2}
\left[
\frac
{a'(s^{\nu_L}_{23}b+c_{23}^{\nu_L}ce^{i(\phi^{\nu}_{e}-\phi^{\nu}_{f})})}{X}
+
\frac
{a(s^{\nu_L}_{23}b'+c_{23}^{\nu_L}c'e^{i(\phi^{\nu}_{e}-\phi^{\nu}_{f})})}{X}
+e^{i\phi_e}
\frac{d\sqrt{|e|^2+|f|^2}}{Y}
\right]
\label{1m13t}
\eeq
Again the leading order $1/Y$ terms have cancelled in
Eq.\ref{1m12t}.

The 13 neutrino angle is given from Eq.\ref{nu13L}
where the leading order form for
$\tilde{m}^{\nu}_{13}$ is given in Eq.\ref{1m13t}, and
for $m_3'$ in Eq.\ref{m3p3}.
In the large $d$ limit we find the same result as in Eq.\ref{2nu13L},
\beq
\theta_{13}^{\nu_L}\approx \frac{|d|}{\sqrt{|e|^2+|f|^2}}
\label{4nu13L}
\eeq
On the other hand in the small $d$ limit we find the leading order result
\beq
\theta_{13}^{\nu_L}\approx\frac{1}{m_3'}e^{-i\phi^{\nu_L}_2}
\left[
\frac{a'(s^{\nu_L}_{23}b+c_{23}^{\nu_L}ce^{i(\phi^{\nu}_{e}-\phi^{\nu}_{f})})}
{X}
+
\frac{a(s^{\nu_L}_{23}b'+c_{23}^{\nu_L}c'
e^{i(\phi^{\nu}_{e}-\phi^{\nu}_{f})})}
{X}
\right]
\label{5nu13L}
\eeq
where the phase $\phi^{\nu_L}_2$ is fixed by
Eq.\ref{absphase}.

After the 13 rotation the neutrino mass matrix is in block diagonal
form as in Eq.\ref{1nustep3}.
Using the leading order form for $\tilde{m}^{\nu_L}_{13}$
in Eq.\ref{1m13t}, and for $m_3'$ in Eq.\ref{m3p3},
and $\tilde{m}^{\nu_L}_{11}$ in Eq.\ref{off2},
we find 
\beq
\tilde{m}^{\nu_L}_{11}\approx 
\frac{2aa'}{X}-e^{-i\phi^{\nu}_{e}}\frac{2d}{\sqrt{|e|^2+|f|^2}}
\left[
\frac{a'(s^{\nu_L}_{23}b+c_{23}^{\nu_L}ce^{i(\phi^{\nu}_{e}-\phi^{\nu}_{f})})}
{X}
+
\frac{a(s^{\nu_L}_{23}b'+c_{23}^{\nu_L}c'
e^{i(\phi^{\nu}_{e}-\phi^{\nu}_{f})})}
{X}
\right]
\label{1m11t}
\eeq
where the leading $1/Y$ dependence in $\tilde{m}^{\nu_L}_{11}$
has cancelled, as before.
After the phase transformations $P_1^{\nu_L}$ the upper block
which remains to be diagonalised is, approximately,
\beq
\left( \begin{array}{cc}
2AA'& A'B+AB'\\
A'B+AB' & 2BB'
\end{array}
\right) 
\label{22block2}
\eeq
where
\beq
A'=\frac{a'}{\sqrt{X}}, \ \ B'= e^{-i(\phi^{\nu_L}_2+\chi^{\nu_L})}
\frac{(c_{23}^{\nu_L}b'-s^{\nu_L}_{23}c'e^{i(\phi^{\nu}_{e}-\phi^{\nu}_{f})})}
{\sqrt{X}}
\label{ApBp}
\eeq
and $A,B$ are as in Eq.\ref{AB}.
The determinant of Eq.\ref{22block2} is now non-zero due to the
two sub-dominant right-handed neutrinos,
\beq
m_1'm_2'\approx 
-(A'B-AB')^2
\label{1det}
\eeq
The trace of Eq.\ref{22block2} is,
\beq
m_1'+m_2'\approx 2(AA'+BB')
\label{1trace}
\eeq
The leading order 
12 neutrino mixing angle is given from Eq.\ref{nu12L},\ref{22block2} 
\beq
\tan 2\theta_{12}^{\nu_L}\approx
\tan \left(\theta_{12}^{\nu_L \rm{seq}} 
+\theta_{12}^{\nu_L '}\right)
\label{2theta12off}
\eeq
where the angle $\theta_{12}^{\nu_L \rm{seq}}$ is given by
\beq
\tan \theta_{12}^{\nu_L \rm{seq}} 
\approx
\frac{A}{B}\approx 
\frac{ae^{i(\phi^{\nu_L}_2+\chi^{\nu_L})}}
{(c_{23}^{\nu_L}b-s^{\nu_L}_{23}ce^{i(\phi^{\nu}_{e}-\phi^{\nu}_{f})})}
\label{4nu12L}
\eeq
and $\theta_{12}^{\nu_L '}$ is given by 
\beq
\tan \theta_{12}^{\nu_L '} 
\approx
\frac{A'}{B'}\approx 
\frac{a'e^{i(\phi^{\nu_L}_2+\chi^{\nu_L})}}
{(c_{23}^{\nu_L}b'-s^{\nu_L}_{23}c'e^{i(\phi^{\nu}_{e}-\phi^{\nu}_{f})})}
\label{5nu12L}
\eeq
The phase $\chi^{\nu_L}$ is fixed by a similar proceedure
to that described previously.

From Eqs.\ref{m3p3},\ref{1det},\ref{1trace},\ref{srhnd2} we see that
off-diagonal sub-dominance has generated a partial neutrino mass hierarchy
\beq
m_1\simlt m_2\ll m_3
\label{partialhi}
\eeq
and hence
\beq
\Delta m_{32}^2\approx m_3^2, \ \  \Delta m_{21}^2=m_2^2-m_1^2.
\label{delta2}
\eeq
Off-diagonal sub-dominance
also naturally leads to the LMA MSW solution
by assuming a mild hierarchy of couplings
in Eq.\ref{srhnd2}.

Analagous to the sequential sub-dominance case,
there is a lower bound on $\theta_{13}^{\nu_L}$ arising from
small $d$ limit limit Eq.\ref{5nu13L}. 
If we assume no accidental cancellations then
the square of the numerator in Eq.\ref{5nu13L} 
is of the same order as the product
of eigenvalues in Eq.\ref{1det}. Since the lightest two eigenvalues
are of similar magnitude in this case, we may deduce a lower
bound analagous to Eq.\ref{13bound2},
\beq
|U_{e3}|^2\simgt |\Delta m_{21}^2/\Delta m_{32}^2|
\label{13bound3}
\eeq

\subsubsection{The pseudo-Dirac limit}
A limiting case of off-diagonal sub-dominance is when
the lightest two neutrinos form an approximate
pseudo-Dirac pair, which happens when the terms in the upper block in
Eq.\ref{22block2} satisfy either
\beq
|A'|,|B|\gg |A|,|B'|
\label{1}
\eeq
or
\beq
|A|,|B'|\gg |A'|,|B|
\label{2}
\eeq
The conditions for this happening are, respectively,
using Eqs.\ref{4nu12L},\ref{5nu12L},
\beq
\tan \theta_{12}^{\nu_L \rm{seq}} \ll 1, \ \ 
\tan \theta_{12}^{\nu_L '} \gg 1,
\eeq
or
\beq
\tan \theta_{12}^{\nu_L \rm{seq}} \gg 1, \ \ 
\tan \theta_{12}^{\nu_L '} \ll 1,
\eeq
In either of the two limiting cases, from Eq.\ref{2theta12off},
we will have an almost maximal neutrino contribution to the solar
mixing angle,
\beq
\theta_{12}^{\nu_L}\approx \pi/4
\eeq
and, since the trace of the upper block is very small,
opposite sign mass eigenvalues,
\beq
m_1'\approx -m_2'
\eeq
both of which are characteristic features of a pseudo-Dirac neutrino pair.
In the pseudo-Dirac case we clearly have
\beq
\Delta m_{21}^2\ll m_2^2 \approx m_1^2\ll m_3^2\approx \Delta m_{32}^2
\label{delta3}
\eeq
The hierarchy in Eq.\ref{delta3} implies that
the pseudo-Dirac case is applicable to the LOW solution
rather then the LMA MSW solution, so is not of interest to us here.

Finally, the 13 neutrino mixing angle
in the pseudo-Dirac case is given in the small $d$ limit from 
Eq.\ref{5nu13L}, using Eqs.\ref{1},\ref{2},
\beq
\theta_{13}^{\nu_L}\approx\frac{m_2'}{m_3'}e^{i\chi^{\nu_L}}
\eeq
Hence in the pseudo-Dirac case we have the bound analagous to 
Eq.\ref{13bound1}
\beq
|\theta_{13}^{\nu_L}|\simgt |m_2/m_3|
\label{13bound4}
\eeq
This bound differs from that quoted in \cite{Lavignac:2002gf},
since it is not singular for $\theta_{12}=\pi/4$.

\subsection{Inverted hierarchy from off-diagonal dominance}

We now discuss the case of off-diagonal right-handed neutrino dominance
which gives rise to a neutrino mass matrix as in Eq.\ref{I}
corresponding to an inverted neutrino mass hierarchy with opposite
sign mass eigenvalues for the heavier pseudo-Dirac neutrino pair. 
The mechanism is based on the off-diagonal form of the right-handed
neutrino mass matrix as in Eq.\ref{off1}, which after the see-saw 
mechanism gives rise to the physical neutrino mass matrix in
Eq.\ref{off2},
\beq
m^{\nu}_{LL}
=
\left( \begin{array}{ccc}
\frac{2aa'}{X}+\frac{d^2}{Y}
& \frac{a'b}{X}+\frac{ab'}{X}+ \frac{de}{Y}
& \frac{a'c}{X}+\frac{ac'}{X}+\frac{df}{Y}    \\
.
& \frac{2bb'}{X}+\frac{e^2}{Y} 
& \frac{b'c}{X}+\frac{bc'}{X}+\frac{ef}{Y}    \\
.
& .
& \frac{2cc'}{X}+\frac{f^2}{Y} 
\end{array}
\right)
\label{off-diagonal1}
\eeq
However, instead of applying the condition for
single right-handed neutrino dominance as in Eq.\ref{srhnd2},
we shall now require the pseudo-Dirac pair of right-handed neutrinos
with mass $X$ to dominate in such a way as to give the leading order
form of the neutrino mass matrix in Eq.\ref{I}.
The two alternative conditions for this are either:
\bea
\frac{|a'b|}{X} & \approx & \frac{|a'c|}{X} \gg 
\frac{|a'a|}{X},\frac{|x'x|}{X},\frac{|yy'|}{Y}
\label{3}
\eea
where $x'\in b',c'$, and $x\in a,b,c$, and $y,y'\in d,e,f$.

Or:
\bea
\frac{|ab'|}{X} & \approx & \frac{|ac'|}{X} \gg 
\frac{|aa'|}{X},\frac{|xx'|}{X},\frac{|yy'|}{Y}
\label{4}
\eea
where $x\in b,c$, and $x'\in a',b',c'$, and $y,y'\in d,e,f$.
In the following we shall consider the first alternative in Eq.\ref{3} for
definiteness.

The proceedure for diagonalising the form of the mass matrix in 
Eq.\ref{off-diagonal1} with the inverted hierarchy conditions in 
Eq.\ref{3} is outlined in appendix \ref{inverted}.
We first perform the re-phasing as in Eq.\ref{mnu2}.
Then we determine the 23 neutrino mixing angle
$\theta_{23}^{\nu_L}$ from Eq.\ref{invnu223L},
\beq
\tan \theta_{23}^{\nu_L}\approx
-e^{i(\phi^{\nu_L}_2-\phi^{\nu_L}_3)}\frac{c}{b}
(1+\tilde{\epsilon_1}-\tilde{\epsilon_2})
\label{i230}
\eeq
where
\beq
\tilde{\epsilon_1}=
\frac{\frac{ac'}{X}+\frac{df}{Y}}{\frac{a'c}{X}},\ \ 
\tilde{\epsilon_2}= \frac{\frac{ab'}{X}+\frac{de}{Y}}{\frac{a'b}{X}}
\eeq
The leading order result, neglecting the corrections
$\tilde{\epsilon_i}$ is then
\beq
\tan \theta_{23}^{\nu_L}\approx
\frac{|c|}{|b|}
\label{i23}
\eeq
where the relative phases are fixed by
\beq
\phi^{\nu_L}_2-\phi^{\nu_L}_3=\pi + \phi^{\nu}_c-\phi^{\nu}_b
\eeq
We find at leading order
\beq
\tilde{m}^{\nu}_{12}=e^{i(\phi^{\nu}_b-\phi^{\nu_L}_2)}
\frac{a'\sqrt{|b|^2+|c|^2}}{X}
\label{im12t}
\eeq
Henceforth we only give the leading order results explicitly,
as in the previous cases. 
The lower block elements are given by
\beq
\tilde{m}^{\nu_L}_{23}\approx
s^{\nu_L}_{23}c^{\nu_L}_{23} 
e^{-i2\phi^{\nu_L}_2}
\left[
\frac{2\sqrt{|b|^2+|c|^2}e^{i\phi^{\nu}_b}
(c^{\nu_L}_{23}b'-s_{23}^{\nu_L}c'e^{i(\phi^{\nu}_{b}-\phi^{\nu}_{c})})}{X}
+
\frac{(e^2-f^2e^{2i(\phi^{\nu}_{b}-\phi^{\nu}_{c})})}
{Y}
\right]
\label{im23t}
\eeq
\beq
\tilde{m}^{\nu_L}_{22}\approx
e^{-i2\phi^{\nu_L}_2}
\left[
\frac{2\sqrt{|b|^2+|c|^2}e^{i\phi^{\nu}_b}
(c^{\nu_L}_{23}b'+s_{23}^{\nu_L}c'e^{i(\phi^{\nu}_{b}-\phi^{\nu}_{c})})}{X}
+
\frac{(c^{\nu_L}_{23}e+s^{\nu_L}_{23}fe^{i(\phi^{\nu}_{b}-\phi^{\nu}_{c})})^2}
{Y}
\right]
\label{im22t}
\eeq
\beq
m'_3\approx
e^{-i2\phi^{\nu_L}_2}
\frac{(s^{\nu_L}_{23}e-c^{\nu_L}_{23}fe^{i(\phi^{\nu}_{b}-\phi^{\nu}_{c})})^2}
{Y}
\label{im3p}
\eeq
We next perform the small angle 13 rotation in Eq.\ref{invnustep3}.
From Eqs.\ref{invnu13L},\ref{im12t},\ref{im23t} we determine
$\theta_{13}^{\nu_L}$,
\beq
\theta_{13}^{\nu_L}\approx -\frac{\tilde{m}^{\nu_L}_{23}}
{\tilde{m}^{\nu}_{12}}
\label{i13}
\eeq
The requirement that $\theta_{13}^{\nu_L}$ is real fixes the absolute
value of the phases $\phi^{\nu_L}_2$, $\phi^{\nu_L}_3$.

After the $P_1^{\nu_L}$ re-phasing the upper $2\times 2$ block which 
remains to be diagonalised is of the form
\beq
\left( 
\begin{array}{cc}
\delta & 1\\
1 & \epsilon
\end{array}
\right)m' 
\label{i22block}
\eeq
where from Eq.\ref{im12t},\ref{im22t},
\beq
m'\approx 
e^{-i\chi^{\nu_L}}
\tilde{m}^{\nu}_{12}
\label{imp}
\eeq
\beq
\delta \approx \frac{1}{m'}(\frac{2aa'}{X}+\frac{d^2}{Y})
\label{idelta}
\eeq
\beq
\epsilon \approx 
e^{-i\chi^{\nu_L}}
\frac{\tilde{m}^{\nu}_{22}}{\tilde{m}^{\nu}_{12}}
\label{iepsilon}
\eeq

Eq.\ref{i22block} is of the pseudo-Dirac form.
The trace and determinant of Eq.\ref{i22block} reveal the two
heavier eigenvalues $m_1',m_2'$,
\beq
m_1'+m_2'=(\delta+\epsilon)m'
\label{itr}
\eeq
\beq
m_1'm_2'=(\delta\epsilon-1)(m')^2
\label{idet}
\eeq
Thus we have
\beq
-m_1'\approx m_2' \approx m'
\label{mp}
\eeq
\beq
\Delta m_{32}^2\approx -|m_2'|^2
\label{i32}
\eeq
\beq
\Delta m_{21}^2\approx 2|m_2'|^2Re(\delta+\epsilon)
\label{i21}
\eeq
We find the neutrino contribution to the 12 mixing angle
by diagonalising Eq.\ref{i22block},
\beq
\tan \theta_{12}^{\nu_L}\approx \frac{2}{\epsilon - \delta}
\label{i12}
\eeq
which is almost maximal due to the smallness of 
$\delta ,\epsilon$ or equivalently the pseudo-Dirac form.

From Eqs.\ref{i21},\ref{i32} we find
\beq
R\equiv \frac{|\Delta m_{21}^2|}{|\Delta m_{32}^2|}\approx 2Re(\delta+\epsilon)
\label{iR}
\eeq
From Eq.\ref{i12} we find
\beq
1-\tan \theta_{12}^{\nu_L}\approx \frac{\epsilon - \delta}{2}
\label{i1-12}
\eeq
These relations together with Eq.\ref{i13} give important constraints
for the inverted hierarchical case. 
By comparing Eq.\ref{iepsilon} to Eq.\ref{i13} it is apparent that,
in the absence of cancellations, 
$\theta_{13}^{\nu_L} \sim \epsilon$.
The parameter $\delta$ is completely 
independent of $\theta_{13}^{\nu_L}$ and $\epsilon$.
Thus for example we can have $\theta_{13}^{\nu_L}=\epsilon =0$,
while at the same time having $\delta \neq 0$ corresponding to 
$R\neq 0$. Thus, unlike the hierarchical case, there is no
lower bound on $\theta_{13}^{\nu_L}$. On the other hand there is 
an upper bound on $\theta_{13}^{\nu_L}$, in the absence of
cancellation effects, which is saturated when $\delta$ is not
large compared to $\epsilon \sim \theta_{13}^{\nu_L}$.
For all choices of $\epsilon \sim \theta_{13}^{\nu_L}$ and $\delta$ we find
\beq
|\theta_{13}^{\nu_L}|\simlt |1-\tan \theta_{12}^{\nu_L}|
\sim R
\label{ibound}
\eeq
Since for the LMA MSW solution we know that $R\sim 10^{-2}$,
Eq.\ref{ibound} implies that the neutrino mixing angles
$\theta_{12}^{\nu_L}=\pi/4$ and $\theta_{13}^{\nu_L}=0$ to high
accuracy. However the LMA MSW solution is not consistent with
maximal mixing and has a preferred mixing angle
$\theta_{12}\approx \pi/6$,
and for this reason there have been claims in the literature
(see for example \cite{Lavignac:2002gf}) that the inverted 
hierarchy case is inconsistent with the LMA MSW solution.
However this conclusion ignores the contribution to MNS angles
arising from the charged lepton sector, as was originally
pointed out in \cite{King:2001ce}. 
It is clear from Eqs.\ref{chlep13},\ref{chlep12}
that even if $\theta_{12}^{\nu_L}=\pi/4$ 
and $\theta_{13}^{\nu_L}=0$ that a reasonable charged lepton contribution,
for example a Cabibbo-like contribution to $\theta_{12}^{E_L}\sim 0.2$,
can result in $\theta_{12}$ in the LMA MSW range and 
$\theta_{13}$ close to the current CHOOZ limit.

\section{Physical applications}

\subsection{Neutrinoless double beta decay}
There has been a recent claim of a signal in neutrinoless double 
beta decay correponding to $|m_{ee}|=0.11-0.56$ eV at 95\% C.L.
However this claim has been criticised by two groups
\cite{Feruglio:2002af}, \cite{Aalseth:2002dt} and in turn this
criticism has been refuted \cite{Klapdor-Kleingrothaus:2002kf}. 
Since the
Heidelberg-Moscow experiment has almost reached its full
sensitivity, we may have to wait for a next generation experiment
such as GENIUS to resolve this question.

From the theoretical point of view, in the natural hierarchical 
and inverted hierarchical models favoured here, corresponding
to the leading order mass matrices in Eqs.\ref{H}, \ref{I},
it is immediately clear that very small values of 
$|m_{ee}|\ll 0.05$ eV are expected, since both matrices
have leading order zeroes in the 11 position. Since we argue that 
these are the only two natural leading order forms, we
can immediately predict that the signal for neutrinoless double 
beta decay should be below the sensitivity of the 
Heidelberg-Moscow experiment. The next question is whether a more
sensitive experiment such as GENIUS is likely to see a signal?

This question has been fully analysed in \cite{Feruglio:2002af},
using the formula
\beq
|m_{ee}|=|\sum_j (U^{MNS}_{ej})^2m_j|
\label{mee1}
\eeq
and the general conclusions are that for the hierarchical models
$|m_{ee}|=0.0005-0.005$ eV while for the inverted hierarchical models
$|m_{ee}|=0.01-0.057$ eV \cite{Feruglio:2002af}.
It is interesting to re-examine these expectations from the point of
view of the equivalent formulation
\beq
|m_{ee}|=|(V^{E_L}m_{LL}^{\nu}{V^{E_L}}^T)_{11}|
\label{mee2}
\eeq
If we neglect the charged lepton contributions then 
$|m_{ee}|$ is just given by $|(m_{LL}^{\nu})_{11}|$
and we can read-off the prediction for this in the 
different models. Of course the contributions from the
charged lepton sector can be very important, as we have 
emphasised, however what can look like
a cancellation conspiracy in the formulation in Eq.\ref{mee1}
can be readily understood in the formulation in Eq.\ref{mee2},
so the alternative formulation can also be quite instructive,
and we use it here as a simple application of our analytic results.
There are three case to consider and we discuss each in turn.

We shall show that in all the cases considered, full hierarchy,
partial hierarchy, and inverted hierarchy, where the hierarchies
result naturally from right-handed neutrino dominance, that 
neutrinoless double beta decay should have a value of about
of $|m_{ee}|\sim 0.007$ eV assuming the LMA MSW solution.
Such a value is below the sensitivity of the Heidelberg-Moscow 
experiment, but is within the range of future proposals
such as the GENIUS experiment.

\vspace{0.25in}
\noindent (i) Full hierarchy from sequential subdominance
\vspace{0.25in}

In this case, ignoring the charged lepton rotations,
from Eq.\ref{seq3} we find
\beq
|m_{ee}|\approx |(m_{LL}^{\nu})_{11}|\approx |\frac{a^2}{X}+\frac{d^2}{Y}|
\label{mee3}
\eeq
Ignoring the contribution from $\frac{d^2}{Y}$ which is always small
even in the large $d$ limit in Eq.\ref{larged}, and using
Eq.\ref{1m2p} we find
\beq
|m_{ee}|\sim m_2\approx \sqrt{\Delta m_{21}^2}\approx 0.007 \rm{eV} 
\label{mee4}
\eeq
which slightly exceeds the upper end of the quoted hierarchical range
\cite{Feruglio:2002af}.

\vspace{0.25in}
\noindent (ii) Partial hierarchy from off-diagonal subdominance
\vspace{0.25in}

Again, ignoring the charged lepton rotations, from Eq.\ref{off2} we have
\beq
|m_{ee}|\approx |(m_{LL}^{\nu})_{11}|\approx |\frac{2aa'}{X}+\frac{d^2}{Y}|
\label{mee5}
\eeq
Again, ignoring the contribution from $\frac{d^2}{Y}$ which is always
small, and using Eq.\ref{1trace} we again find
\beq
|m_{ee}|\sim m_2\approx \sqrt{\Delta m_{21}^2}\approx 0.007 \rm{eV} 
\label{mee6}
\eeq

In the pseudo-Dirac limit the term $\frac{2aa'}{X}$ would be very
small, so much smaller values of $|m_{ee}|$ would be expected.
However, as we discussed earlier, the pseudo-Dirac limit is not
relevant for the LMA MSW solution.

\vspace{0.25in}
\noindent (iii) Inverted hierarchy from off-diagonal dominance
\vspace{0.25in}

Ignoring the charged lepton rotations as before we find, from
Eq.\ref{off-diagonal1},
\beq
|m_{ee}|\approx |(m_{LL}^{\nu})_{11}|\approx |\frac{2aa'}{X}+\frac{d^2}{Y}|
\label{mee7}
\eeq
Using the results of the previous section with Eq.\ref{mee7} we find 
\beq
|m_{ee}|\sim |\delta . m_2 |\simlt |Rm_2| \simlt 0.01m_2 
\simlt 0.01\sqrt{\Delta m_{32}^2} \simlt 0.0005 \rm{eV} 
\label{mee8}
\eeq
The result in Eq.\ref{mee8} neglects the effect
of charged lepton mixing angles. However we saw in the 
previous section that
the inverted hierarchy case in inconsistent with the LMA MSW solution
unless the effect of charged leptons is considered. Therefore
one would expect the result in Eq.\ref{mee8} to be considerably
affected by charged lepton mixing angles in this case. 
As an example let us consider the effect of the 
12 charged lepton mixing angle on the value of $|m_{ee}|$.
From Eq.\ref{mee2} we find a contribution from $\theta_{12}^{E_L}$,
\beq
|m_{ee}|\approx |\frac{\theta_{12}^{E_L}}{\sqrt{2}}|\sqrt{|\Delta m_{32}^2|}
\sim 
|\frac{\theta_{12}^{E_L}}{\theta_C}|0.008 \rm{eV} 
\label{mee9}
\eeq
where we have scaled the charged lepton angle 
by the Cabibbo angle $\theta_C$.
Thus, if the inverted hierarchy case is to achieve consistency with
the LMA MSW solution, we would expect a contribution to 
neutrinoless double beta decay to be given by Eq.\ref{mee9}.

\subsection{Leptogenesis}

Leptogenesis \cite{yanagida1}, \cite{luty}
provides another example where a correct treatment
of phases is crucial to the physics, and this provides another
example of the application of our general treatment including phases
considered in this paper. An analytic treatment of leptogenesis
in the the framework of single right-handed
neutrino dominance was considered earlier \cite{Hirsch:2001dg}. 
In this subsection we highlight how the more general treatment
of phases considered here impacts on our previous analytic estimates.
The numerical results previously presented \cite{Hirsch:2001dg}
are of course completely unchanged.

According to leptogenesis the baryon asymmetry of the universe
is given by
\beq
Y_B=\frac{\alpha}{\alpha -1}d\frac{\epsilon_1}{g^*}
\eeq
where $\alpha$ is a number of order unity arising from
sphaleron effects which convert the lepton number to baryon number,
$d$ is a dilution factor which is crucial to the success of
leptogenesis since it can be many orders of magnitude below unity
if the right-handed neutrinos are not produced efficiently
enough or if they do not satisfy the decay-out-of-equilibrium 
condition strongly enough, and $g^*\sim 10^2$ represents the number of 
effective degrees of freedom. In the analytic discussion here we
shall confine our attention to $\epsilon_1$ which is the
CP decay asymmetry parameter of the lightest right-handed
neutrino, and is the basic seed of lepton number violation
in the leptogenesis scenario.
\beq
\epsilon_1 = \frac{\Gamma ( {N}_{R1} \rightarrow {L_j} +
    H_2 ) - \Gamma({N}^\dagger_{R1} \rightarrow
    {L_j}^\dagger + H^\dagger_2)}
{\Gamma( {N}_{R1} \rightarrow {L_j} +
    H_2) + \Gamma({N}^\dagger_{R1} \rightarrow
    {L_j}^\dagger + H^\dagger_2) }
\label{cp}
\eeq
Assuming a hierarchy of right-handed neutrino masses
$M_1\ll M_2\ll M_3$, one has, approximately,
\beq
\epsilon_1 \approx -\frac{3}{16\pi((Y^{\nu}_{LR})^\dagger Y^{\nu}_{LR})_{11}}
\sum_{i\neq 1} 
Im \left(  \left[ ((Y^{\nu}_{LR})^\dagger Y^{\nu}_{LR})_{1i}\right]^2 \right) 
\left(\frac{M_{1}}{M_{i}} \right)
\label{BP}
\eeq
Leptogenesis therefore depends on following Yukawa combinations,
or, equivalently, Dirac mass combinations
\beq
((Y^{\nu}_{LR})^\dagger Y^{\nu}_{LR})_{1i}
=
\frac{1}{v_2^2}((m^{\nu}_{LR})^\dagger m^{\nu}_{LR})_{1i}
\label{YdagY}
\eeq
The combinations in Eq.\ref{YdagY} have the property that they
are invariant under a change of charged lepton basis,
so they may be calculated either in the original theory basis
or in the diagonal charged lepton mass basis. However the combinations
are sensitive to a basis change of the right-handed neutrinos.
For the present discussion we shall assume that in the theory
basis the right-handed neutrinos are diagonal as in Eq.\ref{diag}.
We shall also assume right-handed neutrino dominance with 
sequential subdominance. Within this class
of model, leptogenesis provides a means of discriminanting
between different subclasses of sequential subdominance,
namely between the case where the dominant right-handed neutrino
is the heaviest ($Y\equiv M_3$) and the case where it is
the lightest ($Y\equiv M_1$), where we assume $M_1\ll M_2\ll M_3$
in both cases. We found that leptogenesis actually prefers
the case $Y\equiv M_3$ where the dominant right-handed neutrino
is the heaviest \cite{Hirsch:2001dg}. The reason is that leptogenesis
is only marginally consistent with the gravitino constraint
on the reheat temperature $T_R\leq 10^9$ GeV, and the values of
$\epsilon_1$ for the case where the dominant right-handed neutrino
is the lighest one are suppressed by a factor of
$m_2/m_3$ relative to the case where it is the heaviest, and this
suppression is enough to make it quantitatively unnacceptable
\cite{Hirsch:2001dg}. We now consider the two possibilities explicitly.

\vspace{0.25in}
\noindent (a) Sequential subdominance with $Y\ll X\ll X'$
\vspace{0.25in}

In this case we need to re-order the matrices involving
right-handed neutrinos so that the first column corresponds to $Y$, and so on.
Then the combinations in Eq.\ref{YdagY} can be written as
\beq
((m^{\nu}_{LR})^\dagger m^{\nu}_{LR})=
\left( \begin{array}{ccc}
<z_1|z_1> & <z_1|z_2> & <z_1|z_3> \\
<z_2|z_1> & <z_2|z_2> & <z_2|z_3> \\
<z_3|z_1> & <z_3|z_2> & <z_3|z_3>
\end{array}
\right) 
\label{z}
\eeq
where we define three complex vectors
$|z_1>=(d,e,f)$, $|z_2>=(a,b,c)$,
$|z_3>=(a',b',c')$.
Clearly the combinations in Eq.\ref{YdagY} only depend on the
first row of the matrix in Eq.\ref{z}, and hence only
on two phases, namely that of $<z_1|z_2>$ and $<z_1|z_3>$.
\footnote{In general a third phase will enter leptogenesis via $<z_2|z_3>$,
but in the approximation that lepton number arises from the
decay of the lightest right-handed neutrino it is not relevant.}
From the sequential subdominance conditions in
Eqs.\ref{srhnd}, we find that one term dominates
\beq
\epsilon_1^{(a)} 
\approx -\frac{3}{16\pi v_2^2}\frac{Y}{X}
\frac{Im \left(<z_1|z_2>^2 \right)}{<z_1|z_1>} 
\eeq
so that leptogenesis only depends on one phase combination.
Using Eq.\ref{def}, and the leading order results
this may be expressed as
\beq
\epsilon_1^{(a)} 
\approx -\frac{3}{16\pi v_2^2}\frac{Y}{X}
Im \left( e^{-2i\phi^{\nu}_{e}}
(s^{\nu_L}_{23}b+c_{23}^{\nu_L}ce^{i(\phi^{\nu}_{e}-\phi^{\nu}_{f})})^2
\right)
\eeq
Using Eqs.\ref{m3p2},\ref{1m2p}
we find the order of magnitude result 
\beq
\epsilon_1^{(a)} 
\sim 2\times 10^{-6}\frac{m_2}{m_3}
\sin (2\phi_{12})\left(\frac{Y}{10^{10}\rm{GeV}}\right)
\label{a}
\eeq
where $\phi_{12}=\arg <z_1|z_2>$,
which quantitatively turns out to be too small due to the $\frac{m_2}{m_3}$
suppression factor \cite{Hirsch:2001dg}.

\vspace{0.25in}
\noindent (b) Sequential subdominance with $X'\ll X\ll Y$
\vspace{0.25in}

The case where the dominant right-handed neutrino is the heaviest
requires no re-ordering of the mass matrices, and we have Eq.\ref{z}
with $|z_1>=(a',b',c')$, $|z_2>=(a,b,c)$,
$|z_3>=(d,e,f)$ (if $M_1=X',M_2=X,M_3=Y$).
From the sequential subdominance conditions in
Eqs.\ref{srhnd}, we find the dominant term is now
\beq
\epsilon_1^{(b)} 
\approx -\frac{3}{16\pi v_2^2}\frac{X'}{Y}
\frac{Im \left(<z_1|z_3>^2 \right)}{<z_1|z_1>} 
\eeq
so that leptogenesis now depends on the $<z_1|z_3>$ phase combination.
Using Eq.\ref{def}, and the leading order results
this may be expressed as
\beq
\epsilon_1^{(b)} 
\approx -\frac{3}{16\pi v_2^2}\frac{X'}{Y}
\frac{|e|^2+|f|^2}{|a'|^2+|b'|^2+|c'|^2}
Im \left( e^{2i\phi^{\nu}_{e}}
(s^{\nu_L}_{23}(b')^{\ast}
+c_{23}^{\nu_L}(c')^{\ast}e^{i(\phi^{\nu}_{f}-\phi^{\nu}_{e})})^2
\right)
\eeq
Using Eqs.\ref{m3p2},
we find the order of magnitude result 
\beq
\epsilon_1^{(b)} 
\sim 2\times 10^{-6}\sin (2\phi_{13})\left(\frac{X'}{10^{10}\rm{GeV}}\right)
\label{b}
\eeq
where $\phi_{13}=\arg <z_1|z_3>$,
which quantitatively turns out to be just about acceptable since
it does not suffer from the $\frac{m_2}{m_3}$
suppression factor of Eq.\ref{a} \cite{Hirsch:2001dg}.

Note that the leptogenesis phase $\phi_{13}$ in this case involves the
phases of the Dirac couplings to the 
lightest right-handed neutrino of mass $X'$.
These couplings and phases are completely irrelevant for constructing
the MNS matrix.

\section{Summary and Conclusion}

The latest solar and atmospheric neutrino data strongly point towards
a minimal interpretation consisting of three active neutrinos
whose mixings are described by a LMA MNS matrix, corresponding to the
mixing angles $\theta_{23} \approx \pi/4$, $\theta_{12} \approx \pi/6$, 
$\theta_{13} \simlt 0.2$. We have argued on
naturalness grounds in favour of a neutrino mass matrix whose
leading order form is
either Eq.\ref{H}, corresponding to a neutrino mass
hierarchy, or Eq.\ref{I} corresponding to an inverse mass hierarchy
with the heavier neutrinos comprising an approximate pseudo-Dirac
pair. We have constructed the LMA MNS matrix
to leading order in $\theta_{13}$ including the neutrino and
charged lepton mixing angles and phases, the latter playing
a crucial r\^{o}le in allowing the inverted hierarchy solution to 
be consistent with the LMA MSW solution. 

We then showed how the neutrino mass matrices with leading order 
forms in Eqs.\ref{H},\ref{I} may be constructed
naturally from the see-saw mechanism with right-handed
neutrino dominance. For the hierarchical
form of the mass matrix in Eq.\ref{H}, single right-handed
neutrino dominance ensures that we obtain a natural 
neutrino mass hierarchy $m_2\ll m_3$ with no tuning,
not even at the level of 10\% accidental cancellations, which would
otherwise be required. Although such a tuning is quite mild,
it would mean that the low energy results would be subject to
rather large radiative corrections, and therefore that the 
low energy neutrino masses and mixing angles are not directly related to the
high energy ones. By contrast in technically natural theories
based on single right-handed neutrino dominance, the radiative
corrections are expected to be quite small, at the level of a few
per cent \cite{King:2000hk}.
For the inverted hierarchy case,
where the leading order form of the mass matrix in Eq.\ref{I}
arises from off-diagonal right-handed neutrino dominance, the
radiative corrections are also small \cite{King:2001ce}.

With right-handed neutrino dominance because the radiative corrections 
are small the low energy observables provide a direct
window into the high energy parameters of the theory.
In such theories it is therefore meaningful to relate the
high energy see-saw parameters to the low energy neutrino observables
by simple analytic relations which ignore the radiative corrections.
We have done this for 
three types of right-handed neutrino dominance corresponding to
a full hierarchy from single right-handed neutrino dominance with
sequential sub-dominance, a partial hierarchy from single right-handed
neutrino dominance with off-diagonal subdominance,
and an inverted mass hierarchy from off-diagonal dominance.
In each case we have derived analytic expressions
for neutrino masses, mixing angles and phases in terms of the
parameters of the see-saw matrices. Our results are accurate
to leading order in $\theta_{13}$, which is sufficiently
accurate bearing in mind that such expressions ignore the effects
of radiative corrections. The leading order expressions are
particularly instructive in guiding the construction of see-saw models,
and these should be particularly useful for unified model building.
This analysis extends the range of the analytic results in
\cite{King:2000mb},\cite{King:2001ce} to include
the important effects of phases and charged lepton mixing angles
which are often overlooked in the literature.

It is worth highlighting how the large atmospheric and solar
mixing angles arise for the different types of
right-handed neutrino dominance. In all cases we write the 
neutrino complex Dirac mass matrix in the common notation
of Eq.\ref{dirac}
where the columns of this matrix represent the couplings to the
different right-handed neutrinos to the right-handed neutrinos
with Majorana mass matrices given in Eqs.\ref{diag},\ref{off-diag}.
Without loss of generality the columns of the matrices may be
permuted corresponding to a re-ordering of right-handed neutrinos.

For case of a full neutrino mass
hierarchy $m_1\ll m_2\ll m_3$
with the right-handed
neutrinos having the texture in Eq.\ref{diag}
with the right-handed neutrino of mass $Y$ giving the
dominant contribution and the right-handed neutrino
of mass $X$ giving the sub-dominant contribution,
and the right-handed neutrino of mass $X'$ giving negligible
contributions, the large atmospheric angle then
arises from Eq.\ref{seq123L},
and the large solar angle arises from Eq.\ref{1nu12L}.
The fact that the solar angle is given, up to phases, by the
ratio of the subdominant coupling $a$ to the difference of subdominant 
couplings $b-c$, was previously known \cite{King:2000mb}.
Eq.\ref{2nu12L} shows that the cancellation in the
denominator in general depends on the relative phases of
$b,c$. In addition the solar angle will receive
corrections from the charged lepton sector
from Eq.\ref{chlep12}.

For the case of a partial neutrino mass hierarchy
$m_1\simlt m_2\ll m_3$ arising from 
the right-handed neutrinos having the texture
in Eq.\ref{off-diag} with the right-handed neutrino 
with mass $Y$ dominating and the off-diagonal
right-handed neutrinos of pseudo-Dirac mass $X$ 
giving the sub-dominant contributions,
the atmospheric angle is determined similarly to the
previous case. However in this case the large solar angle 
arises from a sum of two angles as in Eq.\ref{2theta12off},
where the angle $\theta_{12}^{\nu_L \rm{seq}}$ is given by
Eq.\ref{4nu12L} while 
the angle $\theta_{12}^{\nu_L '}$ is given by Eq.\ref{5nu12L}.
In this case a large solar angle can result from
one or both of the angles $\theta_{12}^{\nu_L \rm{seq}}$,
$\theta_{12}^{\nu_L '}$ being large.
As before cancellations can occur in the
denominators, which in general depends on the relative phases of
$b,c$ and $b',c'$ and in addition the solar angle will receive
corrections from the charged lepton sector
from Eq.\ref{chlep12}. The pseudo-Dirac limit is not relevant for the
LMA MSW solution. For both the full and partial hierarchy cases
we obtain a limit $|U_{e3}|\simgt 0.1$, just below the current CHOOZ
limit. 

For the case of an inverted neutrino mass hierarchy with
a heavier approximately pseudo-Dirac neutrino pair, 
$-m_1\approx m_2\gg m_3$, arising from 
the right-handed neutrinos having the texture
in Eq.\ref{off-diag}, with now the off-diagonal pseudo-Dirac right-handed
neutrinos of mass $X$ dominating, the large atmospheric angle
arises from Eq.\ref{i23},
assuming the condition in Eq.\ref{3},
similar to \cite{King:2001ce}. In this case the 
contribution to the solar angle from the neutrino sector
is almost maximal and we have the result in Eq.\ref{ibound}.
However, including the effect of charged lepton mixing angles,
for example a Cabibbo-like contribution to $\theta_{12}^{E_L}\sim 0.2$,
can result in $\theta_{12}$ in the LMA MSW range and 
$\theta_{13}$ close to the current CHOOZ limit,
as in the hierarchical case.

Although we only quote the leading order
expressions, they should serve to provide a useful guide
to unified model building. More accurate expressions including the 
$m_2/m_3$ corrections can readily be obtained if required from
our more general results.

We have also considered two physical applications of our results,
to neutrinoless double beta decay and to leptogenesis.
Both hierarchical and inverted hierarchical cases predict small
$\beta \beta_{0\nu}$ with $|m_{ee}|\sim 0.007$ eV within
the sensitivity of future proposals such as GENIUS.
We also considered leptogenesis for the
sequential sub-dominance case and emphasised that
successful leptogenesis is possible if the dominant right-handed
neutrino is the heaviest one, but the leptogenesis phase is
unrelated to the MNS phases. Although the leptogenesis phase
is not directly related to the MNS phases, it should be emphasised
that they all originate from the same source, namely
the Dirac mass matrix. It is natural to assume that
the neutrino Dirac mass matrix is complex, in which case
we would expect both a non-zero leptogenesis phase and non-zero MNS
phases. The observation of the Dirac MNS phase at a neutrino factory
would therefore be at least circumstantial evidence for a non-zero
leptogenesis phase.

To conclude, recent SNO results when combined with other
solar and atmosheric neutrino data supports the LMA MSW solution 
and three active neutrinos.
The application of the notion of technical naturalness then
leads to two possible mass patterns for neutrinos, either a hierarchy
or an inverted hierarchy with an approximately pseudo-Dirac heavier
neutrino pair. Further naturalness requirements applied to the
see-saw mechanism, leads us to suppose that the successful mass
patterns result from right-handed neutrino dominance where one
or two right-handed neutrinos dominate and give the leading order
mass matrices in Eqs.\ref{H},\ref{I}. The maximal atmospheric
angle results from a ratio of dominant right-handed neutrino 
Dirac couplings. The large solar angle results
from a ratio of subdominant right-handed neutrino couplings.
The leading order results we present here should provide
a useful and reliable guide in constructing unified models.
The usefulness is due to the simplicity of the leading order results
and the reliability is due to the correct treatment of phases
and also the fact that the radiative corrections are expected 
to be relatively small due to the technical naturalness.
Both these features are a consequence of right-handed neutrino 
dominance, which may be tested experimentally via its predictions
of $\theta_{13}\simgt 0.1$ and $|m_{ee}|\sim 0.007$, assuming the
LMA MSW solution to be correct, although we emphasise that 
both predictions are rather soft since they rely on the assumption
of the absence of cancellations.

\acknowledgments
S.K. would like to thank M.Hirsch for useful conversations and
PPARC for a Senior Fellowship.  

\appendix
\section{Equivalence of different parametrisations}
\label{equivalence}

In this appendix we exhibit the equivalence of different
parametrisations of the MNS matrix.
A $3\times 3$ unitary matrix may be parametrised by 3 angles and 6
phases. We shall find it convenient to parametrise a unitary
matrix $V^{\dagger}$ by
\footnote{It is convenient to define the
parametrisation of $V^{\dagger}$ rather than $V$
because the MNS matrix involves ${V^{\nu_L}}^{\dagger}$
and the neutrino mixing angles will play a central r\^{o}le.}:
\beq
V^{\dagger}=P_2R_{23}R_{13}P_1R_{12}P_3
\label{V1}
\eeq
where $R_{ij}$ are a sequence of real rotations corresponding to the 
Euler angles $\theta_{ij}$, and $P_i$ are diagonal phase matrices.
The Euler matrices are given by
\begin{equation}
R_{23}=
\left(\begin{array}{ccc}
1 & 0 & 0 \\
0 & c_{23} & s_{23} \\
0 & -s_{23} & c_{23} \\
\end{array}\right)
\label{R23}
\end{equation}
\begin{equation}
R_{13}=
\left(\begin{array}{ccc}
c_{13} & 0 & s_{13} \\
0 & 1 & 0 \\
-s_{13} & 0 & c_{13} \\
\end{array}\right)
\label{R13}
\end{equation}
\begin{equation}
R_{12}=
\left(\begin{array}{ccc}
c_{12} & s_{12} & 0 \\
-s_{12} & c_{12} & 0\\
0 & 0 & 1 \\
\end{array}\right)
\label{R12}
\end{equation}
where $c_{ij} = \cos\theta_{ij}$ and $s_{ij} = \sin\theta_{ij}$. 
The phase matrices are given by
\beq
P_1=
\left( \begin{array}{ccc}
1 & 0 & 0    \\
0 & e^{i\chi} & 0 \\
0 & 0 & 1
\end{array}
\right) 
\label{P1}
\eeq
\beq
P_2=
\left( \begin{array}{ccc}
1 & 0 & 0    \\
0 & e^{i\phi_2} & 0 \\
0 & 0 & e^{i\phi_3}
\end{array}
\right) 
\label{P2}
\eeq
\beq
P_3=
\left( \begin{array}{ccc}
e^{i\omega_1} & 0 & 0    \\
0 & e^{i\omega_2} & 0 \\
0 & 0 & e^{i\omega_3}
\end{array}
\right) 
\label{P3}
\eeq

By commuting the phase matrices to the left, it is not difficult to
show that the parametrisation in Eq.\ref{V1} is equivalent to
\beq
V^{\dagger}=PU_{23}U_{13}U_{12}
\label{V2}
\eeq
where $P=P_1P_2P_3$ and
\begin{equation}
U_{23}=
\left(\begin{array}{ccc}
1 & 0 & 0 \\
0 & c_{23} & s_{23}e^{-i\delta_{23}} \\
0 & -s_{23}e^{i\delta_{23}} & c_{23} \\
\end{array}\right)
\label{U23}
\end{equation}
\begin{equation}
U_{13}=
\left(\begin{array}{ccc}
c_{13} & 0 & s_{13}e^{-i\delta_{13}} \\
0 & 1 & 0 \\
-s_{13}e^{i\delta_{13}} & 0 & c_{13} \\
\end{array}\right)
\label{U13}
\end{equation}
\begin{equation}
U_{12}=
\left(\begin{array}{ccc}
c_{12} & s_{12}e^{-i\delta_{12}} & 0 \\
-s_{12}e^{i\delta_{12}} & c_{12} & 0\\
0 & 0 & 1 \\
\end{array}\right)
\label{U12}
\end{equation}
where
\beq
\delta_{23}=\chi+\omega_2-\omega_3
\eeq
\beq
\delta_{13}=\omega_1-\omega_3
\eeq
\beq
\delta_{12}=\omega_1-\omega_2
\eeq
The matrix $U_{MNS}$ is an example of a unitary matrix, 
and as such
it may be parametrised by either of the equivalent forms 
in Eqs.\ref{V1} or \ref{V2}. 
If we use the form in Eq.\ref{V2} then the phase matrix $P$ on the
left may always be removed by an additional charged lepton phase rotation 
$\Delta V^{E_L}=P^\dagger$, 
which is always possible since 
right-handed charged lepton phase rotations can always make the charged
lepton masses real. Therefore $U_{MNS}$ can always be parametrised by
\beq
U_{MNS}=U_{23}U_{13}U_{12}
\label{MNS2A}
\eeq
which involves just three irremoveable physical phases $\delta_{ij}$.
In this parametrisation the Dirac phase $\delta$ 
which enters the CP odd part of
neutrino oscillation probabilities is given by
\beq
\delta = \delta_{13}-\delta_{23}-\delta_{12}.
\label{DiracA}
\eeq

Another common parametrisation of the MNS matrix is
\beq
U_{MNS}=R_{23}U_{13}R_{12}P_0
\label{MNS3}
\eeq
where 
\beq
P_0=
\left( \begin{array}{ccc}
e^{i\beta_1} & 0 & 0    \\
0 & e^{i\beta_2} & 0 \\
0 & 0 & 1
\end{array}
\right) 
\eeq
and in Eq.\ref{MNS3} $U_{13}$ is of the form in
Eq.\ref{U13} but with $\delta_{13}$ replaced by the Dirac phase $\delta$.
The parametrisation in Eq.\ref{MNS3} can be transformed into
the parametrisation in Eq.\ref{MNS2A} by commuting the phase matrix $P_0$
in Eq.\ref{MNS3} to the left, and then removing the phases
on the left-hand side by charged lepton phase rotations.
The two parametrisations are then related by the phase relations
\beq
\delta_{23}=\beta_2
\eeq
\beq
\delta_{13}=\delta + \beta_1
\eeq
\beq
\delta_{12}=\beta_1-\beta_2
\eeq
The use of the parametrisation in Eq.\ref{MNS3} is widespread in the
literature, however for the reasons discussed in the next sub-section 
we prefer to use the parametrisation in Eq.\ref{MNS2A}
which is trivially related to Eq.\ref{MNS3} by the above phase
relations.

\section{Charged lepton contributions to the MNS matrix}
\label{charged}

In this appendix we discuss the contribution of the
charged lepton mixing angles to the MNS matrix.
The MNS matrix is constructed in Eq.\ref{MNS} as a product
of a unitary matrix from the charged lepton sector $V^{E_L}$
and a unitary matrix from the neutrino sector ${V^{\nu_L}}^{\dagger}$.
Each of these unitary matrices may be parametrised by the
parametrisation of $V^{\dagger}$ in Eq.\ref{V1}.
Thus we write
\beq
{V^{\nu_L}}^{\dagger}
=P_2^{\nu_L}R_{23}^{\nu_L}R_{13}^{\nu_L}P_1^{\nu_L}R_{12}^{\nu_L}P_3^{\nu_L}
\label{VnuLB}
\eeq
\beq
{V^{E_L}}^{\dagger}
=P_2^{E_L}R_{23}^{E_L}R_{13}^{E_L}P_1^{E_L}R_{12}^{E_L}P_3^{E_L}
\label{VELB}
\eeq
where the Euler angles and phases are defined as in 
Eqs.\ref{R23}-\ref{P3} 
but now there are independent angles
and phases for the left-handed neutrino and charged lepton sectors
distinguished by the superscripts $\nu_L$ and $E_L$.
The MNS matrix from Eqs.\ref{MNS},\ref{VnuLB},\ref{VELB} is then
\beq
U_{MNS}=
{P_3^{E_L}}^{\dagger}{R_{12}^{E_L}}^{\dagger}{P_1^{E_L}}^{\dagger}
{R_{13}^{E_L}}^{\dagger}{R_{23}^{E_L}}^{\dagger}{P_2^{E_L}}^{\dagger}
P_2^{\nu_L}R_{23}^{\nu_L}R_{13}^{\nu_L}P_1^{\nu_L}R_{12}^{\nu_L}P_3^{\nu_L}
\label{MNS4}
\eeq
As before we commute all the phase matrices to the left, then choose
${P_3^{E_L}}^{\dagger}$ to cancel all the phases on the left-hand
side, to leave just
\beq
U_{MNS}=
{U_{12}^{E_L}}^{\dagger}
{U_{13}^{E_L}}^{\dagger}{U_{23}^{E_L}}^{\dagger}
U_{23}^{\nu_L}U_{13}^{\nu_L}U_{12}^{\nu_L}
\label{MNS5}
\eeq
with independent phases and angles 
for the left-handed neutrino and charged lepton sectors,
in the convention of Eqs.\ref{U23},\ref{U13},\ref{U12}.
The phases in Eq.\ref{MNS5} are given in terms of the phases in
Eqs.\ref{VnuLB}, \ref{VELB} by
\bea
\delta_{12}^{\nu_L}&=&\omega_1^{\nu_L}-\omega_2^{\nu_L}
\\
\delta_{13}^{\nu_L}&=&\omega_1^{\nu_L}-\omega_3^{\nu_L}
\\
\delta_{23}^{\nu_L}&=&\chi^{\nu_L}+\omega_2^{\nu_L}-\omega_3^{\nu_L}
\\
\delta_{23}^{E_L}&=&
-\phi_2^{E_L}+\phi_3^{E_L}+\phi_2^{\nu_L}-\phi_3^{\nu_L}
+\chi^{\nu_L}+\omega_2^{\nu_L}-\omega_3^{\nu_L}
\\
\delta_{13}^{E_L}&=&\phi_3^{E_L}-\phi_3^{\nu_L}
+\omega_1^{\nu_L}-\omega_3^{\nu_L}
\\
\delta_{12}^{E_L}&=&\chi^{E_L}+\phi_2^{E_L}-\phi_2^{\nu_L}
-\chi^{\nu_L}+\omega_1^{\nu_L}-\omega_2^{\nu_L}
\eea
The form of $U_{MNS}$ in Eq.\ref{MNS5} is similar to the 
parametrisation in Eq.\ref{MNS2}, which is the practical reason
why we prefer that form rather than that in Eq.\ref{MNS3}.

We now discuss the MNS matrix to leading order
in $\theta_{13}$. 
From Eqs.\ref{MNS2A},\ref{U23},\ref{U13},\ref{U12}, we find to leading
order in $\theta_{13}$ that $U_{MNS}$ may be expanded as:
\bea
&&U_{MNS} \approx   \nonumber \\
&& \left(\begin{array}{ccc}
c_{12} & s_{12}e^{-i\delta_{12}} & \theta_{13}e^{-i\delta_{13}} \\
-s_{12}c_{23}e^{i\delta_{12}}-c_{12}s_{23}\theta_{13}
e^{i(\delta_{13}-\delta_{23})}
& c_{12}c_{23}-s_{12}s_{23}\theta_{13}
e^{i(-\delta_{23}+\delta_{13}-\delta_{12})}
& s_{23}e^{-i\delta_{23}} \\
s_{12}s_{23}e^{i(\delta_{23}+\delta_{12})}
-c_{12}c_{23}\theta_{13}e^{i\delta_{13}}
& -c_{12}s_{23}e^{i\delta_{23}}
-s_{12}c_{23}\theta_{13}e^{i(\delta_{13}-\delta_{12})}
& c_{23} \\
\end{array}\right) \nonumber \\
\label{MNS6}
\eea
For $\theta_{13} = 0.1$, close to the CHOOZ limit,
the approximate form in Eq.\ref{MNS6} is accurate to 1\%.

We now wish to expand the MNS matrix in terms of neutrino
and charged lepton mixing angles and phases
to leading order in small angles,
using Eq.\ref{MNS5}. 
In technically natural theories, based on right-handed
neutrino dominance, the contribution to $\theta_{23}$ comes mainly
from the neutrino sector, $\theta_{23}\approx \theta_{23}^{\nu_L}$.
Furthermore in natural theories we expect
that the contributions to $\theta_{13}$
are all separately small so that the smallness of this angle
does not rely on accidental cancellations.
Clearly this implies that $\theta_{13}^{\nu_L}$ and $\theta_{13}^{E_L}$
must both be $\simlt \theta_{13}$. Since the 13 element
of $U_{MNS}$ also receives a contribution from the charged lepton
sector proportional to $s_{12}^{E_L}s_{23}^{\nu_L}$, the 
same argument also implies that 
$\theta_{12}^{E_L}\simlt \theta_{13}$. 
Therefore the natural expectation is that all the charged lepton
mixing angles are small! Expanding Eq.\ref{MNS5} to leading order
in small angles $\theta_{12}^{E_L}$, 
$\theta_{23}^{E_L}$, $\theta_{13}^{E_L}$, $\theta_{13}^{\nu_L}$, we find
\bea
&&U_{MNS} \approx   \nonumber \\
&& \left(\begin{array}{ccc}
c_{12}^{\nu_L} & s_{12}^{\nu_L}e^{-i\delta_{12}^{\nu_L}} & 
\theta_{13}^{\nu_L}e^{-i\delta_{13}^{\nu_L}} \\
-s_{12}^{\nu_L}c_{23}^{\nu_L}e^{i\delta_{12}^{\nu_L}}
-c_{12}^{\nu_L}s_{23}^{\nu_L}\theta_{13}^{\nu_L}
e^{i(\delta_{13}^{\nu_L}-\delta_{23}^{\nu_L})}
& c_{12}^{\nu_L}c_{23}^{\nu_L}-s_{12}^{\nu_L}s_{23}^{\nu_L}\theta_{13}^{\nu_L}
e^{i(-\delta_{23}^{\nu_L}+\delta_{13}^{\nu_L}-\delta_{12}^{\nu_L})}
& s_{23}^{\nu_L}e^{-i\delta_{23}^{\nu_L}} \\
s_{12}^{\nu_L}s_{23}^{\nu_L}e^{i(\delta_{23}^{\nu_L}+\delta_{12}^{\nu_L})}
-c_{12}^{\nu_L}c_{23}^{\nu_L}\theta_{13}^{\nu_L}e^{i\delta_{13}^{\nu_L}}
& -c_{12}^{\nu_L}s_{23}^{\nu_L}e^{i\delta_{23}^{\nu_L}}
-s_{12}^{\nu_L}c_{23}^{\nu_L}\theta_{13}^{\nu_L}
e^{i(\delta_{13}^{\nu_L}-\delta_{12}^{\nu_L})}
& c_{23}^{\nu_L} \nonumber \\
\end{array}\right) \nonumber \\
&&+\theta_{23}^{E_L} 
\left(\begin{array}{ccc}
c_{12}^{\nu_L} &  s_{12}^{\nu_L}e^{-i\delta_{12}^{\nu_L}} & 0
\nonumber \\
-s_{23}^{\nu_L}s_{12}^{\nu_L}
e^{i(\delta_{23}^{\nu_L}-\delta_{23}^{E_L}
+\delta_{12}^{\nu_L})} &
s_{23}^{\nu_L}c_{12}^{\nu_L}
e^{i(\delta_{23}^{\nu_L}-\delta_{23}^{E_L})} &
-c_{23}^{\nu_L}e^{-i\delta_{23}^{E_L}}
\nonumber \\
-c_{23}^{\nu_L}s_{12}^{\nu_L}
e^{i(\delta_{23}^{E_L}+\delta_{12}^{\nu_L})} &
c_{23}^{\nu_L}c_{12}^{\nu_L}e^{i\delta_{23}^{E_L}} &
s_{23}^{\nu_L}e^{i(\delta_{23}^{E_L}-\delta_{23}^{\nu_L})} 
\nonumber \\
\end{array}\right) \nonumber \\
&&+\theta_{13}^{E_L}
\left(\begin{array}{ccc}
-s_{12}^{\nu_L}s_{23}^{\nu_L}
e^{i(\delta_{12}^{\nu_L}+\delta_{23}^{\nu_L}-\delta_{13}^{E_L})} &  
c_{12}^{\nu_L}s_{23}^{\nu_L}e^{i(\delta_{23}^{\nu_L}-\delta_{13}^{E_L})} &  
-c_{23}^{\nu_L}e^{-i\delta_{13}^{E_L}} \nonumber \\
0 & 0 & 0 \nonumber \\
c_{12}^{\nu_L}e^{i\delta_{13}^{E_L}} 
& s_{12}^{\nu_L}e^{i(-\delta_{12}^{\nu_L}+\delta_{13}^{E_L})} & 0 
\nonumber \\
\end{array}\right) \nonumber \\
&&+\theta_{12}^{E_L}
\left(\begin{array}{ccc}
c_{23}^{\nu_L}s_{12}^{\nu_L}e^{i(\delta_{12}^{\nu_L}-\delta_{12}^{E_L})} &  
-c_{23}^{\nu_L}c_{12}^{\nu_L}e^{-i\delta_{12}^{E_L}} &  
s_{23}^{\nu_L}e^{i(-\delta_{23}^{\nu_L}-\delta_{12}^{E_L})} \nonumber \\
c_{12}^{\nu_L}e^{i\delta_{12}^{E_L}} &
s_{12}^{\nu_L}e^{i(-\delta_{12}^{\nu_L}+\delta_{12}^{E_L})} & 0 \nonumber \\
0 & 0 & 0  \\
\end{array}\right) \\
\label{MNS7}
\eea
where we have dropped terms of order $\theta_{23}^{E_L}\theta_{13}$.
The first matrix on the right hand side of
Eq.\ref{MNS7} gives the contribution to the
MNS matrix from the neutrino mixing angles and phases, and is of 
the same form as Eq.\ref{MNS6}.
The subsequent matrices give the corrections to the MNS matrix
from the charged lepton mixing angles 
$\theta_{23}^{E_L}$, $\theta_{13}^{E_L}$, and $\theta_{12}^{E_L}$.

\section{Proceedure for diagonalising hierarchical mass matrices}  
\label{proceedure}
In this appendix we discuss the diagonalisation of 
a general complex hierarchical matrix $m$ where
\beq
m=
\left( \begin{array}{ccc}
m_{11} & m_{12} & m_{13} \\
m_{21} & m_{22} & m_{23} \\
m_{31} & m_{32} & m_{33}
\end{array}
\right) 
\label{m1}
\eeq
The matrix $m$ is diagonalised by a sequence of tranformations:
\beq
{P_3^{L}}^{\ast}{R_{12}^{L}}^{T}{P_1^{L}}^{\ast}
{R_{13}^{L}}^{T}{R_{23}^{L}}^{T}{P_2^{L}}^{\ast}
m
P_2^{R}R_{23}^{R}R_{13}^{R}P_1^{R}R_{12}^{R}P_3^{R}=
\left( \begin{array}{ccc}
m_1 & 0 & 0    \\
0 & m_2 & 0 \\
0 & 0 & m_3
\end{array}
\right) 
\label{diag5}
\eeq
In the case of the charged lepton mass matrix, all the rotation
angles are small, while in the case of the neutrino mass matrix it is
symmetric. The results for the general complex matrix
$m$ will be sufficiently general to allow us to 
apply them to both of the physical cases of interest as limiting cases.

The proceedure for diagonalising a general hierarchical matrix $m$
involves the following steps.

1. The first step involves multiplying the mass matrix $m$ by the inner
phase matrices $P_2$ defined in Eq.\ref{P2}:
\beq
{P_2^{L}}^{\ast}
m
P_2^{R}=
\left( \begin{array}{ccc}
m_{11} & m_{12}e^{i\phi_2^R} & m_{13}e^{i\phi_3^R} \\
m_{21}e^{-i\phi_2^L} & m_{22}e^{i(\phi_2^R-\phi_2^L)} 
& m_{23}e^{i(\phi_3^R-\phi_2^L)} \\
m_{31}e^{-i\phi_3^L} 
& m_{32}e^{i(\phi_2^R-\phi_3^L)} 
& m_{33}e^{i(\phi_3^R-\phi_3^L)}
\end{array}
\right) 
\equiv
\left( \begin{array}{ccc}
m_{11} & m_{12}'& m_{13}' \\
m_{21}' & m_{22}' & m_{23}' \\
m_{31}' & m_{32}' & m_{33}'
\end{array}
\right) 
\label{step1}
\eeq
The purpose of this re-phasing is to facilitate steps 2,3 using
real rotation angles $\theta_{23}$, $\theta_{13}$, as we shall see.

2. The second step is to perform the real rotations $R_{23}$ 
defined in Eq.\ref{R23} on the re-phased matrix from step1.
The purpose is to put zeroes in the 23,32 elements of the 
resulting matrix:
\beq
{R_{23}^{L}}^{T}
\left( \begin{array}{ccc}
m_{11} & m_{12}'& m_{13}' \\
m_{21}' & m_{22}' & m_{23}' \\
m_{31}' & m_{32}' & m_{33}'
\end{array}
\right) 
R_{23}^{R}
\equiv
\left( \begin{array}{ccc}
m_{11} & \tilde{m}_{12} & \tilde{m}_{13} \\
\tilde{m}_{21} &  \tilde{m}_{22} &  0 \\
\tilde{m}_{31} &  0 &  m_3'
\end{array}
\right) 
\label{step2}
\eeq
The zeroes in the 23,32 positions are achieved by diagonalising the
lower 23 block, using the reduced matrix $R_{23}$ obtained by striking
out the row and column in which the unit element appears, to leave 
a $2\times 2$ rotation,
\beq
{R_{23}^{L}}^{T}
\left( \begin{array}{cc}
m_{22}' & m_{23}' \\
m_{32}' & m_{33}'
\end{array}
\right) 
R_{23}^{R}
\equiv
\left( \begin{array}{cc}
\tilde{m}_{22} &  0 \\
0 &  m_3'
\end{array}
\right) 
\label{step21}
\eeq
which implies
\beq
\tan 2\theta_{23}^L=
\frac{2\left[m_{33}'m_{23}'+m_{22}'m_{32}'\right]}
{\left[{m_{33}'}^2-{m_{22}'}^2
+{m_{32}'}^2-{m_{23}'}^2\right]}
\label{23L}
\eeq
\beq
\tan 2\theta_{23}^R=
\frac{2\left[m_{33}'m_{32}'
+m_{22}'m_{23}'\right]}
{\left[{m_{33}'}^2-{m_{22}'}^2
+{m_{23}'}^2-{m_{32}'}^2\right]}
\label{23R}
\eeq
The requirement that the angles $\theta_{23}^L$ and $\theta_{23}^R$
are real means that the numerators and denominators must have equal 
phases, and this is achieved by adjusting the relative phases
$\phi_i^R-\phi_j^L$ which appear in the lower block of Eq.\ref{step1}.
The remaining elements are then given by the reduced rotations
\beq
\left( \begin{array}{cc}
\tilde{m}_{12} & \tilde{m}_{13}
\end{array}
\right) 
=
\left( \begin{array}{cc}
m_{12}'& m_{13}'
\end{array}
\right) 
R_{23}^{R}
\label{step22}
\eeq
\beq
\left( \begin{array}{c}
\tilde{m}_{21} \\
\tilde{m}_{31} 
\end{array}
\right) 
=
{R_{23}^{L}}^{T}
\left( \begin{array}{c}
m_{21}' \\
m_{31}' 
\end{array}
\right) 
\label{step23}
\eeq

3. The third step is to perform the real small angle rotations $R_{13}$ 
defined in Eq.\ref{R13} on the matrix from step 2.
The purpose is to put zeroes in the 13,31 elements of the 
resulting matrix:
\beq
{R_{13}^{L}}^{T}
\left( \begin{array}{ccc}
m_{11} & \tilde{m}_{12} & \tilde{m}_{13} \\
\tilde{m}_{21} &  \tilde{m}_{22} &  0 \\
\tilde{m}_{31} &  0 &  m_3'
\end{array}
\right) 
R_{13}^{R}
\approx
\left( \begin{array}{ccc}
\tilde{m}_{11} & \tilde{m}_{12} & 0 \\
\tilde{m}_{21} &  \tilde{m}_{22} &  0 \\
0 &  0 &  m_3'
\end{array}
\right) 
\label{step3}
\eeq
The zeroes in the 13,31 positions are achieved by diagonalising the
outer 13 block, using the reduced matrix $R_{13}$ obtained by striking
out the row and column in which the unit element appears, to leave 
a $2\times 2$ rotation,
\beq
{R_{13}^{L}}^{T}
\left( \begin{array}{cc}
m_{11} & \tilde{m}_{13} \\
\tilde{m}_{31} & m_3'
\end{array}
\right) 
R_{13}^{R}
\approx
\left( \begin{array}{cc}
\tilde{m}_{11} &  0 \\
0 &  m_3'
\end{array}
\right) 
\label{step31}
\eeq
which implies
\beq
\theta_{13}^L\approx
\frac{\tilde{m}_{13}}{m_3'}+
\frac{\tilde{m}_{31}m_{11}}{(m_3')^2}
\label{13L}
\eeq
\beq
\theta_{13}^R\approx
\frac{\tilde{m}_{31}}{m_3'}+
\frac{\tilde{m}_{13}m_{11}}{(m_3')^2}
\label{13R}
\eeq
The requirement that the angles $\theta_{13}^L$ and $\theta_{31}^R$
are real fixes the absolute value of the phases 
$\phi_i^R+\phi_j^L$, since only the relative phases were fixed
previously. This uses up all the phase freedom and thus
all the resulting mass matrix elements in Eq.\ref{step3} remain complex.
Note that Eq.\ref{step3} is written
to leading order in the small angles $\theta_{13}$,
and as discussed previously the 23,32 elements remain zero to this
order. The large complex element $m_3'$ is
approximately unchanged to this order.
Due to the zeroes in the 23,32 position of the matrix
the elements $\tilde{m}_{12}$ and $\tilde{m}_{21}$ are also unchanged
to leading order. The element $\tilde{m}_{22}$ is also unchanged
of course since it is not present in the reduced matrix.
The only new element is therefore
\beq
\tilde{m}_{11}\approx m_{11}-
\frac{\tilde{m}_{13}\tilde{m}_{31}}{m_3'}
\label{m11}
\eeq

4. The fourth step involves multiplying the mass matrix 
resulting from Eq.\ref{step3} by the
phase matrices $P_1$ defined in Eq.\ref{P1}:
\beq
{P_1^{L}}^{\ast}
\left( \begin{array}{ccc}
\tilde{m}_{11} & \tilde{m}_{12} & 0 \\
\tilde{m}_{21} &  \tilde{m}_{22} &  0 \\
0 &  0 &  m_3'
\end{array}
\right) 
P_1^{R}=
\left( \begin{array}{ccc}
\tilde{m}_{11} & \tilde{m}_{12}e^{i\chi^R} & 0 \\
\tilde{m}_{21}e^{-i\chi^L} &  \tilde{m}_{22}e^{i(\chi^R-\chi^L)} &  0 \\
0 &  0 &  m_3'
\end{array}
\right) 
\equiv
\left( \begin{array}{ccc}
\tilde{m}_{11} & \tilde{m}_{12}'& 0 \\
\tilde{m}_{21}' & \tilde{m}_{22}' & 0 \\
0 & 0 & m_3'
\end{array}
\right) 
\label{step4}
\eeq
The purpose of this re-phasing is to facilitate step 5 using
real rotation angle $\theta_{12}$.

5. The fifth step is to perform the real rotations $R_{12}$ 
defined in Eq.\ref{R12} on the re-phased matrix from step 4.
The purpose is to put zeroes in the 12,21 elements of the 
resulting matrix:
\beq
{R_{12}^{L}}^{T}
\left( \begin{array}{ccc}
\tilde{m}_{11} & \tilde{m}_{12}'& 0 \\
\tilde{m}_{21}' & \tilde{m}_{22}' & 0 \\
0 & 0 & m_3'
\end{array}
\right) 
R_{12}^{R}
\equiv
\left( \begin{array}{ccc}
m_1' & 0 & 0 \\
0 & m_2' &  0 \\
0 &  0 & m_3'
\end{array}
\right) 
\label{step5}
\eeq
The zeroes in the 12,21 positions are achieved by diagonalising the
upper 12 block, using the reduced matrix $R_{12}$ obtained by striking
out the row and column in which the unit element appears, to leave 
a $2\times 2$ rotation,
\beq
{R_{12}^{L}}^{T}
\left( \begin{array}{cc}
\tilde{m}_{11} & \tilde{m}_{12}'\\
\tilde{m}_{21}' & \tilde{m}_{22}'
\end{array}
\right) 
R_{12}^{R}
\equiv
\left( \begin{array}{cc}
m_1' & 0 \\
0 & m_2' 
\end{array}
\right) 
\label{step51}
\eeq
which implies
\beq
\tan 2\theta_{12}^L=
\frac{2\left[\tilde{m}_{22}'\tilde{m}_{12}'
+\tilde{m}_{11}\tilde{m}_{21}'\right]}
{\left[(\tilde{m}_{22}')^2-(\tilde{m}_{11})^2
+(\tilde{m}_{21}')^2-(\tilde{m}_{12}')^2\right]}
\label{12L}
\eeq
\beq
\tan 2\theta_{12}^R=
\frac{2\left[\tilde{m}_{22}'\tilde{m}_{21}'
+\tilde{m}_{11}\tilde{m}_{12}'\right]}
{\left[(\tilde{m}_{22}')^2-(\tilde{m}_{11})^2
+(\tilde{m}_{12}')^2-(\tilde{m}_{21}')^2\right]}
\label{12R}
\eeq
The requirement that the angles $\theta_{12}^L$ and $\theta_{21}^R$
are real means that the numerators and denominators must have equal 
phases, and this is achieved by adjusting the phases $\chi_L$, $\chi_R$.

6. The sixth step involves multiplying the complex diagonal mass matrix 
resulting from Eq.\ref{step5} by the
phase matrices $P_3$ defined in Eq.\ref{P3}:
\beq
{P_3^{L}}^{\ast}
\left( \begin{array}{ccc}
m_1' & 0 & 0 \\
0 & m_2' &  0 \\
0 &  0 & m_3'
\end{array}
\right) 
P_3^{R}=
\left( \begin{array}{ccc}
m_1 & 0 & 0 \\
0 & m_2 &  0 \\
0 &  0 & m_3
\end{array}
\right) 
\label{step6}
\eeq
The result of this re-phasing is a diagonal matrix with real
eigenvalues. In the case of charged leptons this last step can be
achieved by a suitable $P_3^R$ for any choice of $P_3^L$.
This freedom in $P_3^L$ enables three phases to be removed from the MNS
matrix.

\section{Diagonalising the hierarchical neutrino mass matrix}
\label{hierarchical}

In this appendix we shall apply the results of 
appendix \ref{proceedure} to the case of the
complex symmetric hierarchical neutrino mass matrix of the 
leading order form as shown in Eq.\ref{H}, which will be
written in full generality as
\beq
m^{\nu}_{LL}=
\left( \begin{array}{ccc}
m^{\nu}_{11} & m^{\nu}_{12} & m^{\nu}_{13} \\
m^{\nu}_{12} & m^{\nu}_{22} & m^{\nu}_{23} \\
m^{\nu}_{13} & m^{\nu}_{23} & m^{\nu}_{33}
\end{array}
\right) 
\equiv
\left( \begin{array}{ccc}
|m^{\nu}_{11}|e^{i\phi^{\nu}_{11}}
& |m^{\nu}_{12}|e^{i\phi^{\nu}_{12}}
& |m^{\nu}_{13}|e^{i\phi^{\nu}_{13}} \\
|m^{\nu}_{12}|e^{i\phi^{\nu}_{12}} 
& |m^{\nu}_{22}|e^{i\phi^{\nu}_{22}} 
& |m^{\nu}_{23}|e^{i\phi^{\nu}_{23}} \\
|m^{\nu}_{13}|e^{i\phi^{\nu}_{13}} 
& |m^{\nu}_{23}|e^{i\phi^{\nu}_{23}} 
& |m^{\nu}_{33}|e^{i\phi^{\nu}_{33}} 
\end{array}
\right) 
\label{mnu1}
\eeq
where it should be remembered that the elements in the 
lower 23 block are larger than the other elements,
as in Eq.\ref{H}.

The proceedure outlined in appendix \ref{hierarchical}
for diagonalising $m^{\nu}_{LL}$ is to 
work our way from the inner transformations to the
outer transformations as follows.
\begin{enumerate}
\item Re-phase $m^{\nu}_{LL}$ using the $P^{\nu_L}_2$.
\item Put zeroes in the 23=32 positions using $R^{\nu_L}_{23}$.
\item Put zeroes in the 13=31 positions using $R^{\nu_L}_{13}$.
\item Re-phase the mass matrix using $P^{\nu_L}_1$.
\item Put zeroes in the 12=21 positions using $R^{\nu_L}_{12}$.
\item Make the diagonal elements real using the $P^{\nu_L}_3$.
\end{enumerate}
If $\theta^{\nu_L}_{13}$ is small, then for the
hierarchical case $m_3 \gg m_2$ this proceedure
will result in an approximately diagonal matrix to 
leading order in $\theta^{\nu_L}_{13}$. 
One might object that
after step 3 the $R^{\nu_L}_{13}$ rotations will
``fill-in'' the zeroes in the 23,32 positions with terms
of order $\theta^{\nu_L}_{13}$ multiplied by 
$m^{\nu_L}_{12},m^{\nu_L}_{13}$. However in the hierarchical case
$m^{\nu_L}_{12},m^{\nu_L}_{13}$ are smaller than $m^{\nu_L}_{33}$ 
by a factor of $\theta^{\nu_L}_{13}$ which 
means that the ``filled-in'' 23,32 entries
are suppressed by a total factor of $(\theta^{\nu_L}_{13})^2$ compared to the
33 element.
This means that after the 5 steps above a hierarchical
matrix will be diagonal
to leading order in $\theta^{\nu_L}_{13}$, as claimed.
For the inverted hierarchical neutrino case
a different proceedure must be followed, as discussed in the
next sub-section. Here we shall systematically diagonalise the
hierarchical neutrino mass matrix in Eq.\ref{mnu1} by following
the above proceedure as follows.

The first step is to re-phase the matrix in Eq.\ref{mnu1}
using ${P_2^{\nu_L}}^{\ast}$ so that
the neutrino mass matrix becomes,
\beq
\left( \begin{array}{ccc}
|m^{\nu}_{11}|e^{i\phi^{\nu}_{11}}
& |m^{\nu}_{12}|e^{i(\phi^{\nu}_{12}-\phi^{\nu_L}_2)} 
& |m^{\nu}_{13}|e^{i(\phi^{\nu}_{13}-\phi^{\nu_L}_3)}  \\
|m^{\nu}_{12}|e^{i(\phi^{\nu}_{12}-\phi^{\nu_L}_2)} 
& |m^{\nu}_{22}|e^{i(\phi^{\nu}_{22}-2\phi^{\nu_L}_2)} 
& |m^{\nu}_{23}|e^{i(\phi^{\nu}_{23}-\phi^{\nu_L}_2-\phi^{\nu_L}_3)} \\
|m^{\nu}_{13}|e^{i(\phi^{\nu}_{13}-\phi^{\nu_L}_3)} 
& |m^{\nu}_{23}|e^{i(\phi^{\nu}_{23}-\phi^{\nu_L}_2-\phi^{\nu_L}_3)} 
& |m^{\nu}_{33}|e^{i(\phi^{\nu}_{33}-2\phi^{\nu_L}_3)} 
\end{array}
\right) 
\label{mnu2}
\eeq
To determine the 23 neutrino mixing angle
$\theta_{23}^{\nu_L}$ we perform a 23 rotation which 
diagonalises the lower 23 block of Eq.\ref{mnu2}.
From Eq.\ref{23L} we find the 23 neutrino mixing angle
$\theta_{23}^{\nu_L}$ as
\beq
\tan 2\theta_{23}^{\nu_L}=
\frac{2\left[|m^{\nu}_{23}|
e^{i(\phi^{\nu}_{23}-\phi^{\nu_L}_2-\phi^{\nu_L}_3)}
\right]}
{\left[|m^{\nu}_{33}|e^{i(\phi^{\nu}_{33}-2\phi^{\nu_L}_3)}
-|m^{\nu}_{22}|e^{i(\phi^{\nu}_{22}-2\phi^{\nu_L}_2)}
\right]}
\label{nu23L}
\eeq
The relative phase $\phi^{\nu_L}_2-\phi^{\nu_L}_3$ is fixed by the 
requirement that the angle $\theta_{23}^{\nu_L}$ in Eq.\ref{nu23L} be
real,
\beq
|m^{\nu}_{33}|
\sin(\phi^{\nu}_{33}-\phi^{\nu}_{23}+\phi^{\nu_L}_2-\phi^{\nu_L}_3)
=
|m^{\nu}_{22}|
\sin(\phi^{\nu}_{22}-\phi^{\nu}_{23}+\phi^{\nu_L}_3-\phi^{\nu_L}_2)
\label{phase1}
\eeq
After the 23 rotation in Eq.\ref{step2},
the neutrino mass matrix in Eq.\ref{mnu2} becomes
\beq
\left( \begin{array}{ccc}
m^{\nu}_{11} & \tilde{m}^{\nu}_{12} & \tilde{m}^{\nu}_{13} \\
\tilde{m}^{\nu}_{12} &  \tilde{m}^{\nu}_{22} &  0 \\
\tilde{m}^{\nu}_{13} &  0 &  m_3'
\end{array}
\right) 
\label{nustep2}
\eeq
The lower block elements are given by
\beq
\left( \begin{array}{cc}
\tilde{m}^{\nu}_{22} &  0\\
0 &  m_3'
\end{array}
\right) 
\equiv
{R^{\nu_L}_{23}}^T
\left( \begin{array}{cc}
|m^{\nu}_{22}|e^{i(\phi^{\nu}_{22}-2\phi^{\nu_L}_2)}  
& |m^{\nu}_{23}|e^{i(\phi^{\nu}_{23}-\phi^{\nu_L}_2-\phi^{\nu_L}_3)}  \\
|m^{\nu}_{23}|e^{i(\phi^{\nu}_{23}-\phi^{\nu_L}_2-\phi^{\nu_L}_3)}
& |m^{\nu}_{33}|e^{i(\phi^{\nu}_{33}-2\phi^{\nu_L}_3)} 
\end{array}
\right) 
R^{\nu_L}_{23}
\label{nustep222}
\eeq
which implies
\bea
\tilde{m}^{\nu}_{22}&=&
(c_{23}^{\nu_L})^2|m^{\nu}_{22}|e^{i(\phi^{\nu}_{22}-2\phi^{\nu_L}_2)}
-2s_{23}^{\nu_L}c_{23}^{\nu_L}|m^{\nu}_{23}|
e^{i(\phi^{\nu}_{23}-\phi^{\nu_L}_2-\phi^{\nu_L}_3)}
+(s_{23}^{\nu_L})^2|m^{\nu}_{33}|e^{i(\phi^{\nu}_{33}-2\phi^{\nu_L}_3)}
\nonumber \\
&&  \label{m22}\\
m_3'&=&
(s_{23}^{\nu_L})^2|m^{\nu}_{22}|e^{i(\phi^{\nu}_{22}-2\phi^{\nu_L}_2)}
+2s_{23}^{\nu_L}c_{23}^{\nu_L}|m^{\nu}_{23}|
e^{i(\phi^{\nu}_{23}-\phi^{\nu_L}_2-\phi^{\nu_L}_3)}
+(c_{23}^{\nu_L})^2|m^{\nu}_{33}|e^{i(\phi^{\nu}_{33}-2\phi^{\nu_L}_3)}
\nonumber \\
&&
\label{m3p}
\eea
and from Eq.\ref{step23}
\beq
\left( \begin{array}{c}
\tilde{m}^{\nu}_{12} \\
\tilde{m}^{\nu}_{13} 
\end{array}
\right) 
=
{R_{23}^{\nu_L}}^{T}
\left( \begin{array}{c}
|m^{\nu}_{12}|
e^{i(\phi^{\nu}_{12}-\phi^{\nu_L}_2)} \\
|m^{\nu}_{13}| 
e^{i(\phi^{\nu}_{13}-\phi^{\nu_L}_3)}
\end{array}
\right) 
\label{nustep23}
\eeq

We now perform a 13 rotation on the neutrino matrix in
Eq.\ref{nustep2} which diagonalises the outer 13 block of Eq.\ref{nustep2} and
determines the 13 neutrino mixing angle
$\theta_{13}^{\nu_L}$. 
From Eq.\ref{13L} we find the 13 neutrino mixing angle
$\theta_{13}^{\nu_L}$ as
\beq
\theta_{13}^{\nu_L}\approx
\frac{\tilde{m}^{\nu}_{13}}{m_3'}
\label{nu13L}
\eeq
The absolute phases $\phi^{\nu_L}_2$, $\phi^{\nu_L}_3$ are fixed by the 
requirement that the angle $\theta_{13}^{\nu_L}$ in Eq.\ref{nu13L} be
real,
\beq
s_{23}^{\nu_L}|m^{\nu}_{12}|
\sin(\phi^{\nu}_{12}-\phi^{\nu_L}_2-\phi_3')
+
c_{23}^{\nu_L}|m^{\nu}_{13}|
\sin(\phi^{\nu}_{13}-\phi^{\nu_L}_3-\phi_3')=0
\label{absphase}
\eeq
After the 13 rotation in Eq.\ref{step3},
Eq.\ref{nustep2} becomes
\beq
\left( \begin{array}{ccc}
\tilde{m}^{\nu}_{11} & \tilde{m}^{\nu}_{12} & 0 \\
\tilde{m}^{\nu}_{12} &  \tilde{m}^{\nu}_{22} &  0 \\
0 &  0 &  m_3'
\end{array}
\right) 
\equiv
\left( \begin{array}{ccc}
|\tilde{m}^{\nu}_{11}|e^{i\tilde{\phi}^{\nu}_{11}}
& |\tilde{m}^{\nu}_{12}|e^{i\tilde{\phi}^{\nu}_{12}} & 0 \\
|\tilde{m}^{\nu}_{12}|e^{i\tilde{\phi}^{\nu}_{12}} 
& |\tilde{m}^{\nu}_{22}|e^{i\tilde{\phi}^{\nu}_{22}} &  0 \\
0 &  0 &  |m_3'|e^{i\phi_3'}
\end{array}
\right) 
\label{nustep3}
\eeq
To leading order in $\theta_{13}^{\nu_L}$ 
the only new element in Eq.\ref{nustep3} is
\beq
\tilde{m}^{\nu_L}_{11}\approx m^{\nu_L}_{11}-
\frac{(\tilde{m}^{\nu_L}_{13})^2}{m_3'}
\label{num11}
\eeq

It only remains to determine the 12 neutrino mixing angle
$\theta_{12}^{\nu_L}$ by diagonalising the upper 12 block
of Eq.\ref{nustep3}. 
From Eq.\ref{12L} we find the 12 neutrino mixing angle
$\theta_{12}^{\nu_L}$ as
\beq
\tan 2\theta_{12}^{\nu_L}=
\frac{2\left[|\tilde{m}^{\nu}_{12}|
e^{i(\tilde{\phi}^{\nu}_{12}-\chi^{\nu_L})}
\right]}
{\left[|\tilde{m}^{\nu}_{22}|
e^{i(\tilde{\phi}^{\nu}_{22}-2\chi^{\nu_L})}
-|\tilde{m}^{\nu}_{11}|
e^{i\tilde{\phi}^{\nu}_{11}}
\right]}
\label{nu12L}
\eeq
The phase $\chi^{\nu_L}$ is fixed by the 
requirement that the angle $\theta_{12}^{\nu_L}$ in Eq.\ref{nu12L} be
real,
\beq
|\tilde{m}^{\nu}_{22}|
\sin(\tilde{\phi}^{\nu}_{22}-\tilde{\phi}^{\nu}_{12}-\chi^{\nu_L})
=
|\tilde{m}^{\nu}_{11}|
\sin(\tilde{\phi}^{\nu}_{11}-\tilde{\phi}^{\nu}_{12}+\chi^{\nu_L})
\label{chi}
\eeq
After the 12 rotation the upper block of the matrix in
Eq.\ref{nustep3} is diagonal and the resulting matrix is
\beq
\left( \begin{array}{ccc}
m_1' & 0 & 0 \\
0 & m_2' &  0 \\
0 &  0 & m_3'
\end{array}
\right) 
\equiv
\left( \begin{array}{ccc}
m_1e^{i{\phi}_1'} & 0 & 0 \\
0 & m_2e^{i{\phi}_2'} & 0 \\
0 &  0 &  m_3e^{i\phi_3'}
\end{array}
\right) 
\label{nustep5}
\eeq
where from Eq.\ref{step51}
\bea   
m_1'&=&
(c_{12}^{\nu_L})^2|\tilde{m}^{\nu}_{11}|e^{i\tilde{\phi}^{\nu}_{11}}
-2s_{12}^{\nu_L}c_{12}^{\nu_L}|\tilde{m}^{\nu}_{12}|
e^{i(\tilde{\phi}^{\nu}_{12}-\chi^{\nu_L})}
+(s_{12}^{\nu_L})^2|\tilde{m}^{\nu}_{22}|
e^{i(\tilde{\phi}^{\nu}_{22}-2\chi^{\nu_L})}
\nonumber \\
&& \\
m_2'&=&
(s_{12}^{\nu_L})^2|\tilde{m}^{\nu}_{11}|e^{i\tilde{\phi}^{\nu}_{11}}
+2s_{12}^{\nu_L}c_{12}^{\nu_L}|\tilde{m}^{\nu}_{12}|
e^{i(\tilde{\phi}^{\nu}_{12}-\chi^{\nu_L})}
+(c_{12}^{\nu_L})^2|\tilde{m}^{\nu}_{22}|
e^{i(\tilde{\phi}^{\nu}_{22}-2\chi^{\nu_L})}
\nonumber \\
&&
\label{m2p}
\eea

It is a simple matter to adjust the
phases $\omega^{\nu_L}_i$ in $P^{\nu_L}_3$
to remove the phases in Eq.\ref{nustep5} and make 
the neutrino masses real,
as in Eq.\ref{step6},
\beq
\omega^{\nu_L}_i=\frac{{\phi}_i'}{2} 
\eeq
This completes the diagonalisation
in Eq.\ref{diag4}.
In the case of neutrino masses, unlike the case of the charged
fermions, there is no left over phase freedom. This is the reason
why the MNS matrix has three more physical phases than the
CKM matrix.

\section{Diagonalising the inverted hierarchical neutrino mass matrix}
\label{inverted}
In this appendix we shall consider the case of the
complex symmetric inverted hierarchical neutrino mass matrix of the 
leading order form as shown in Eq.\ref{I}.
In this case the proceedure is as follows.

\begin{enumerate}
\item Re-phase $m^{\nu}_{LL}$ using the $P^{\nu_L}_2$.
\item Put zeroes in the 13=31 positions using $R^{\nu_L}_{23}$.
\item Put zeroes in the 23=32 positions using $R^{\nu_L}_{13}$.
\item Re-phase the mass matrix using $P^{\nu_L}_1$.
\item Put zeroes in the 12=21 positions using $R^{\nu_L}_{12}$.
\item Make the diagonal elements real using the $P^{\nu_L}_3$.
\end{enumerate}

We continue to write the neutrino mass matrix as
in Eq.\ref{mnu1}, but now it should be remembered that the 12,13 elements
are now larger than the other elements, as in Eq.\ref{I}.
This is the reason why the above
proceedure differs from that for the case of the hierarchical neutrino
mass matrix. 

We first perform the re-phasing as in Eq.\ref{mnu2}.
Then we determine the 23 neutrino mixing angle
$\theta_{23}^{\nu_L}$ by performing a 23 rotation such that
\beq
\left( \begin{array}{ccc}
m^{\nu}_{11} & \tilde{m}^{\nu}_{12} & 0 \\
\tilde{m}^{\nu}_{12} &  \tilde{m}^{\nu}_{22} &  \tilde{m}^{\nu}_{23}\\
0 &  \tilde{m}^{\nu}_{23} &  m_3'
\end{array}
\right) 
\equiv
{R^{\nu_L}_{23}}^T
\left( \begin{array}{ccc}
|m^{\nu}_{11}|e^{i\phi^{\nu}_{11}}
& |m^{\nu}_{12}|e^{i(\phi^{\nu}_{12}-\phi^{\nu_L}_2)} 
& |m^{\nu}_{13}|e^{i(\phi^{\nu}_{13}-\phi^{\nu_L}_3)}  \\
|m^{\nu}_{12}|e^{i(\phi^{\nu}_{12}-\phi^{\nu_L}_2)} 
& |m^{\nu}_{22}|e^{i(\phi^{\nu}_{22}-2\phi^{\nu_L}_2)} 
& |m^{\nu}_{23}|e^{i(\phi^{\nu}_{23}-\phi^{\nu_L}_2-\phi^{\nu_L}_3)} \\
|m^{\nu}_{13}|e^{i(\phi^{\nu}_{13}-\phi^{\nu_L}_3)} 
& |m^{\nu}_{23}|e^{i(\phi^{\nu}_{23}-\phi^{\nu_L}_2-\phi^{\nu_L}_3)} 
& |m^{\nu}_{33}|e^{i(\phi^{\nu}_{33}-2\phi^{\nu_L}_3)} 
\end{array}
\right) 
R^{\nu_L}_{23}
\label{invnustep2}
\eeq
where
\beq
\left( \begin{array}{c}
\tilde{m}^{\nu}_{12} \\
0 
\end{array}
\right) 
=
{R_{23}^{\nu_L}}^{T}
\left( \begin{array}{c}
|m^{\nu}_{12}|
e^{i(\phi^{\nu}_{12}-\phi^{\nu_L}_2)} \\
|m^{\nu}_{13}| 
e^{i(\phi^{\nu}_{13}-\phi^{\nu_L}_3)}
\end{array}
\right) 
\label{invnustep21}
\eeq
which gives the 23 neutrino mixing angle
$\theta_{23}^{\nu_L}$ in this case to be
\beq
\tan \theta_{23}^{\nu_L}=
\frac{-|m^{\nu}_{13}|
e^{i(\phi^{\nu}_{13}-\phi^{\nu_L}_3)}}
{|m^{\nu}_{12}|e^{i(\phi^{\nu}_{12}-\phi^{\nu_L}_2)}}
\label{invnu23L}
\eeq
Since the Euler angles are constrained to satisfy 
$\theta_{ij}\leq \pi/2$, we must have
$\tan \theta_{23}^{\nu_L} \approx +1$,
and this then fixes 
\beq
\phi^{\nu}_{13}-\phi^{\nu}_{12}+\phi^{\nu_L}_2-\phi^{\nu_L}_3=\pi
\eeq
This fixes $\phi^{\nu_L}_2-\phi^{\nu_L}_3$ and gives
\beq
\tan \theta_{23}^{\nu_L}=
\frac{|m^{\nu}_{13}|}
{|m^{\nu}_{12}|}
\label{invnu223L}
\eeq
and
\beq
\tilde{m}^{\nu}_{12}=c_{23}^{\nu_L}|m^{\nu}_{12}|
e^{i(\phi^{\nu}_{12}-\phi^{\nu_L}_2)}-
s_{23}^{\nu_L}|m^{\nu}_{13}|
e^{i(\phi^{\nu}_{13}-\phi^{\nu_L}_3)}
\eeq
The lower block elements are given by
\beq
\left( \begin{array}{cc}
\tilde{m}^{\nu}_{22} &  \tilde{m}^{\nu}_{23}\\
\tilde{m}^{\nu}_{23} &  m_3'
\end{array}
\right) 
\equiv
{R^{\nu_L}_{23}}^T
\left( \begin{array}{cc}
|m^{\nu}_{22}|e^{i(\phi^{\nu}_{22}-2\phi^{\nu_L}_2)}  
& |m^{\nu}_{23}|e^{i(\phi^{\nu}_{23}-\phi^{\nu_L}_2-\phi^{\nu_L}_3)}  \\
|m^{\nu}_{23}|e^{i(\phi^{\nu}_{23}-\phi^{\nu_L}_2-\phi^{\nu_L}_3)}
& |m^{\nu}_{33}|e^{i(\phi^{\nu}_{33}-2\phi^{\nu_L}_3)} 
\end{array}
\right) 
R^{\nu_L}_{23}
\label{invnustep22}
\eeq
which implies
\beq
\tilde{m}^{\nu}_{23}=
s_{23}^{\nu_L}c_{23}^{\nu_L}
(|m^{\nu}_{22}|e^{i(\phi^{\nu}_{22}-2\phi^{\nu_L}_2)}-
|m^{\nu}_{33}|e^{i(\phi^{\nu}_{33}-2\phi^{\nu_L}_3)})
+((c_{23}^{\nu_L})^2-(s_{23}^{\nu_L})^2)
|m^{\nu}_{23}|
e^{i(\phi^{\nu}_{23}-\phi^{\nu_L}_2-\phi^{\nu_L}_3)}
\label{mt23}
\eeq
and the remaining diagonal elements are given as before in 
Eqs.\ref{m22},\ref{m3p}.

We next perform a small angle 13 rotation such that
\beq
\left( \begin{array}{ccc}
m^{\nu}_{11} & \tilde{m}^{\nu}_{12} & 0 \\
\tilde{m}^{\nu}_{12} &  \tilde{m}^{\nu}_{22} &  0\\
0 &  0 &  m_3'
\end{array}
\right) 
\approx
{R^{\nu_L}_{13}}^T
\left( \begin{array}{ccc}
m^{\nu}_{11} & \tilde{m}^{\nu}_{12} & 0 \\
\tilde{m}^{\nu}_{12} &  \tilde{m}^{\nu}_{22} &  \tilde{m}^{\nu}_{23}\\
0 &  \tilde{m}^{\nu}_{23} &  m_3'
\end{array}
\right) 
R^{\nu_L}_{13}
\label{invnustep3}
\eeq
where
\beq
\left( \begin{array}{c}
\tilde{m}^{\nu}_{12} \\
0 
\end{array}
\right) 
\approx
{R_{13}^{\nu_L}}^{T}
\left( \begin{array}{c}
\tilde{m}^{\nu}_{12}   \\
\tilde{m}^{\nu}_{23}
\end{array}
\right) 
\label{invnustep31}
\eeq
Note that to leading order in $\theta_{13}^{\nu_L}$ the
large element $\tilde{m}^{\nu}_{12}$ is unchanged.
The remaining elements in Eq.\ref{invnustep3} are also unchanged
to leading order in $\theta_{13}^{\nu_L}$.
The 13=31 element in Eq.\ref{invnustep3} gets filled in
by a term $\theta_{13}^{\nu_L}(m^{\nu}_{11}-m_3')$ which is
of order $(\theta_{13}^{\nu_L})^2$ compared to $\tilde{m}^{\nu}_{12}$
and does not appear to leading order in $\theta_{13}^{\nu_L}$.
From Eq.\ref{invnustep31} the 13 neutrino mixing angle
$\theta_{13}^{\nu_L}$ is
\beq
\theta_{13}^{\nu_L}\approx
\frac{-\tilde{m}^{\nu}_{23}}{\tilde{m}^{\nu}_{12}}
\label{invnu13L}
\eeq
The requirement that $\theta_{13}^{\nu_L}$ is real fixes the absolute
value of the phases $\phi^{\nu_L}_2$, $\phi^{\nu_L}_3$.

The left hand side of Eq.\ref{invnustep3} now resembles the left hand
side of Eq.\ref{nustep3}, except that here $m^{\nu}_{11}$ is unchanged
due to the zero 13=31 element after the 23 rotation.
Therefore the rest of the diagonalisation process follows that of
the previous hierarchical case from Eq.\ref{nu12L} onwards,
where now
\beq
\tan 2\theta_{12}^{\nu_L}=
\frac{2\left[|\tilde{m}^{\nu}_{12}|
e^{i(\tilde{\phi}^{\nu}_{12}-\chi^{\nu_L})}
\right]}
{\left[|\tilde{m}^{\nu}_{22}|
e^{i(\tilde{\phi}^{\nu}_{22}-2\chi^{\nu_L})}
-|{m}^{\nu}_{11}|
e^{i{\phi}^{\nu}_{11}}
\right]}
\label{invnu12L}
\eeq
Note that in the inverted hierarchy case here we have
\beq
|\tilde{m}^{\nu}_{12}| \gg |\tilde{m}^{\nu}_{22}|,|{m}^{\nu}_{11}|
\eeq
which implies an almost degenerate pair of
pseudo-Dirac neutrino masses (with opposite sign eigenvalues) and an almost
maximal 12 mixing angle from Eq.\ref{invnu12L}.

\end{document}